\newcommand{\PRE}[1]{}       
\documentclass[aps,prd,preprint,preprintnumbers,superscriptaddress,tightenlines,
amsmath,amssymb,showpacs,nofootinbib]{revtex4-1}
\newcommand{\ssection}[1]{{\em #1.\ }}
\usepackage{amsfonts}
\usepackage{enumerate}
\usepackage{url}
\usepackage{color}
\usepackage{xcolor}
\usepackage{slashed}
\usepackage{subcaption}
\usepackage{booktabs}
\usepackage{simplewick}
\usepackage{bm}
\newcommand{\overbar}[1]{\mkern 1.8mu\overline{\mkern-1.8mu#1\mkern-1.8mu}\mkern 1.8mu}
\usepackage[pdftex]{graphicx}
\usepackage[colorlinks=true,citecolor=blue,urlcolor=blue]{hyperref}
\usepackage[utf8]{inputenc} 

\renewcommand{\eqref}[1]{Eq.~(\ref{#1})}

\begin{document}

\title{\PRE{\vspace*{1.5in}}
Light scalars with lepton number to solve the $(g-2)_e$ anomaly
\PRE{\vspace*{.5in}}}

\author{Susan Gardner\thanks{gardner@pa.uky.edu}}
\affiliation{Department of Physics and Astronomy, University of 
Kentucky, Lexington, Kentucky 40506-0055 USA
\PRE{\vspace*{.2in}}}
\author{Xinshuai Yan\thanks{xinshuai.yan@uky.edu}}
\affiliation{Department of Physics and Astronomy, University of 
Kentucky, Lexington, Kentucky 40506-0055 USA
\PRE{\vspace*{.2in}}}



\begin{abstract} 
\PRE{\vspace*{.2in}} 
Scalars that carry lepton number 
can help mediate would-be 
lepton-number-violating processes, 
such as neutrinoless double $\beta$ decay or 
lepton-scattering-mediated nucleon-antinucleon conversion. 
Here we show that such new scalars can also solve the anomaly in 
precision determinations of the fine-structure constant $\alpha$ from atom
interferometry and from the electron's anomalous magnetic moment, $a_e \equiv (g-2)_e/2$, 
by reducing $|a_e|$. 
Study of the phenomenological constraints on these solutions favor a 
doubly-charged scalar with mass below the GeV scale. 
Significant constraints arise 
from the 
measurement of the parity-violating asymmetry in M{\o}ller scattering, and we consider the
implications of 
the next-generation MOLLER experiment at Jefferson Laboratory
and of an improved $a_e$ measurement. 
\end{abstract}


\maketitle

\section{Introduction}
\label{sec:intro}

Through tour-de-force efforts in both theory and experiment, the 
anomalous magnetic moments of both the electron and muon have emerged as 
exquisitely sensitive probes of physics 
beyond the Standard Model (SM)~\cite{Aoyama2019,Hanneke:2008tm,Bouchendira2011,Parker:2018vye,Bennett:2006fi,Roberts:2010zz,Tanabashi:2018oca,Mohr:2015ccw}. 
For many years, the measured value of the electron's 
anomalous magnetic moment $a_e \equiv (g-2)_e/2$\footnote{In this paper
we define the magnetic moment of a charged lepton $\ell$ as 
$\bm{\mu}_\ell = g_\ell \mathbf{S} e/2m_\ell $, with 
$g_\ell >0$ and $e = - |e|$~\cite{Peskin:1995ev}.} 
was used 
to determine the most precise value of the fine-structure 
constant 
$\alpha\equiv e^2/4\pi \epsilon_0 \hbar c$~\cite{Mohr:2015ccw},
with the measurement of $a_\mu$ providing sensitivity to new physics 
at the weak scale, once the hadronic and electroweak contributions were 
taken into account~\cite{Roberts:2010zz,Lindner:2016bgg}. 
In recent years, with the emergence of precise assessments 
of $a_{e}$ in QED perturbation theory through fifth-order 
in $\alpha/\pi$~\cite{Aoyama:2012wj,Aoyama:2014sxa,Laporta:2017okg,Aoyama_alpha5,Aoyama2019}
and 
precise determinations of $\alpha$~\cite{Bouchendira2011,Parker:2018vye} from 
atom interferometry~\cite{Cronin2009,2019CRPhy..20...77C}, 
$a_e$ itself, due to it quantum nature, has 
also emerged as a probe of physics beyond the SM. Indeed the comparison 
of $a_e$ from its direct experimental measurement with its expected 
value in the SM, using atom interferometry to fix $\alpha$, yields
the most precise test of the SM in all of physics~\cite{Gabrielse:2019cgf}.

The SM value of $a_e$ is dominated by the contribution from 
QED --- though contributions from the SM weak gauge bosons $W^\pm, Z^0$ and hadronic
effects also exist, these are known to be extraordinarily small, 
contributing only 0.026 ppb~\cite{Czarnecki:1995sz,Knecht:2002hr,Czarnecki:2002nt} 
and 1.47 ppb~\cite{Nomura:2012sb,Jegerlehner:2017zsb}, 
respectively, of the total contribution 
to $a_e^{\rm SM}$~\cite{Aoyama2019}. The analysis of atom 
interferometry measurements for $\alpha$ also require the use of QED 
theory and other observables~\cite{Mohr:2015ccw}, though the uncertainty in 
the determined value of $\alpha$ is dominated by that in its measured observable, $h/M_X$, where 
$h$ is Planck's constant and $M_X$ is the mass of atomic species $X$. 
With the most precise experimental result for $a_e$~\cite{Hanneke:2008tm,Hanneke:2010au} and 
$h/M_X$ measurements 
for Rb~\cite{Bouchendira2011} 
 or Cs atoms~\cite{Parker:2018vye} 
to determine 
$\alpha$ and thus $a_e^{\rm SM}$~\cite{Aoyama2019}
we report~\cite{Aoyama2019}
\begin{eqnarray}
a_e^{\rm EXP} - a_e^{\rm SM}\,[{\rm Rb}] &=& (-131 \pm 77)\times 10^{-14}\,, \\
a_e^{\rm EXP} - a_e^{\rm SM}\,[{\rm Cs}] &=& (-88 \pm 36)\times 10^{-14} \,, 
\end{eqnarray}
where here and elsewhere the uncertainties are added in quadrature. 
In what follows we use the most precise determination of $\alpha$ to define the $a_e$ anomaly, 
$\Delta a_e \equiv (-88 \pm 36)\times 10^{-14}$~\cite{Parker:2018vye,Hanneke:2008tm,Aoyama2019},
which is a discrepancy of $\sim 2.4\sigma$. 
For reference we report the anomaly in 
$(g-2)_\mu$ as well~\cite{Blum:2018mom,Bennett:2006fi}
\begin{equation}
\Delta a_\mu \equiv a_\mu^{\rm EXP} - a_\mu^{\rm SM} = (2.74\pm 0.73)\times 10^{-9} \,, 
\end{equation}
for a discrepancy of $\sim 3.7\sigma$, as also determined by 
Ref.~\cite{Keshavarzi:2018mgv}, with a sign opposite to that of 
$\Delta a_e$. Both the relative sign and size of the anomalies suggest distinct 
mechanisms for their explanation. For example, if weak-scale new physics were to explain 
$\Delta a_\mu$, scaling as $m_\mu^2$, then its contribution to $\Delta a_e$ would
be roughly 10 times too small, 
$\Delta a_e \simeq 0.7 \times 10^{-13}$~\cite{Giudice:2012ms,Liu:2018xkx}. 
Thus explaining both anomalies is seemingly not possible in models that differentiate 
electrons and muons only by their mass --- rather, possible solutions should break
lepton flavor 
universality~\cite{Davoudiasl:2018fbb,Crivellin:2018qmi,Liu:2018xkx,Han:2018znu}. 
The relatively large size of $\Delta a_e$ 
also suggests the appearance of new physics below the $\sim 1$ GeV scale. 

Several models of light new physics have been 
proposed that could explain both of the 
$a_\ell$ 
anomalies~\cite{Davoudiasl:2018fbb,Crivellin:2018qmi,Liu:2018xkx,Han:2018znu}. 
However, the suggestion that new physics at scales below $\sim 1$ GeV, arising from 
so-called dark, hidden, or secluded sectors, 
could explain the $a_\mu$ anomaly has existed much 
longer~\cite{Fayet:2007ua,Pospelov2009}. 
Keen interest in such scenarios has been generated not only 
by anomalies in high-energy astrophysics that could arise from 
dark-matter annihilation~\cite{ArkaniHamed:2008qn}, but also 
by an appreciation of the great reaches of untested parameter space possible for their
realization~\cite{Bjorken:2009mm,Batell:2009di}, which has spurred new experimental 
initatives~\cite{Essig:2013lka,Alexander:2016aln}. Although the 
possibility that a U(1) gauge boson that mixes with the photon~\cite{Holdom:1985ag}, 
a ``dark photon,'' 
could explain $\Delta a_\mu$ has been 
ruled out~\cite{Lees:2017lec}, solutions 
involving a new light scalar or pseudoscalar are 
still possible~\cite{Chen:2015vqy,Liu:2016qwd}. 
Since the dark photon gives a positive contribution to $a_\ell$~\cite{Fayet:2007ua,Pospelov2009}, 
it also cannot address the $a_e$ anomaly~\cite{Parker:2018vye}. 

Models that address both $a_\ell$ anomalies treat the electrons and muons in different ways. 
In Ref.~\cite{Davoudiasl:2018fbb}, a single real scalar is introduced, and, in the electron
case, the scalar coupling to a heavy charged fermion, such as the $\tau$, can be chosen to 
mediate a two-loop 
Barr-Zee~\cite{Barr:1990vd} contribution to $a_e$ that yields the needed opposite sign. 
In Ref.~\cite{Crivellin:2018qmi}, models with an abelian flavor symmetry
${\rm L}_{\mu} - {\rm L}_{\tau}$ 
are used to realize different contributions to $a_e$ and $a_\mu$, 
with  the suggested consequence that the permanent
electric-dipole moment (EDM) of the $\mu$ 
could be much larger than supposed from electron EDM limits. 
In Ref.~\cite{Liu:2018xkx}, a complex scalar is introduced with CP-odd couplings
to the electron and CP-even couplings to the muon, generating contributions to $a_{e,\mu}$ of
opposite sign. The somewhat disjoint nature of the 
various simultaneous solutions, and the severity of the constraint from nonobservation of
$\mu\to e \gamma$~\cite{Crivellin:2018qmi}, 
suggests that we can address one
anomaly without precluding the other. In this paper we show that we can 
solve the $a_e$ anomaly by introducing a scalar with lepton number that couples 
to first-generation fermions only, respecting SM symmetries, supposing that one of the solutions 
for $\Delta a_\mu$ 
proposed in 
Refs.~\cite{Kinoshita:1990aj,Barger:2010aj,TuckerSmith:2010ra,Chen:2015vqy,Liu:2016qwd,Batell:2016ove,Liu:2018xkx}, e.g.,  
could also act. The solutions we have found also serve as
ingredients in 
minimal scalar 
models~\cite{Davies:1990sc,Bowes:1996xy,Arnold:2012sd,Arnold:2013cva,Gardner:2018azu} 
that can also 
mediate lepton-number violating processes, 
such as neutrinoless double $\beta$ decay and scattering-mediated 
nucleon-antinucleon conversion~\cite{Gardner:2018azu}.

Giudice, Paradisi, and Passera have shown that many possible new physics models
could generate a shift of $a_e$ from its SM value~\cite{Giudice:2012ms}, 
considering both models that connect to a change in $a_\mu$ by $(m_\mu/m_e)^2$ and
those that do not. In the latter class they consider models that 
connect to violations of charged lepton flavor number or lepton 
flavor universality, as well as models with heavy vector-like 
fermions~\cite{Giudice:2012ms,Kannike:2011ng}. In the last example, Giudice et al. 
introduced a SU(2) vector-like 
doublet and singlet, with 
interactions that can explicitly break lepton number. 
In what follows we consider a new physics model 
for $\Delta a_e$ of a completely different 
kind --- here the scalars carry lepton number, with scalar-fermion interactions
that conserve lepton number, and indeed are SM-gauge invariant; 
and these features are essential to the results 
we find. Other models pertinent to $a_e$~\cite{Girrbach:2009uy,Dutta:2018fge}
that also address the $\Delta a_e$ anomaly~\cite{Dutta:2018fge}
have been proposed. 
Interestingly, models with a new axial-vector boson also generate
contributions that decrease $|a_e|$~\cite{Fayet:2007ua,Kahn:2016vjr}, 
though other empirical constraints exist on these solutions 
as well~\cite{Kahn:2016vjr,Parker:2018vye}. 

Scalars that carry lepton number also appear in neutrino mass models. 
Although the smallness of the neutrino masses can be elegantly ascribed 
to a seesaw mechanism 
with a new-physics scale of 
some $M_{\rm N} \sim 10^{10-15}\,{\rm GeV}$~\cite{Minkowski,GellMann:1980vs,Yanagida:1979as,Mohapatra:1979ia}, 
there are many alternate possibilities~\cite{Gouvea:2016shl}. 
In type II seesaw 
models~\cite{Konetschny:1977bn,Magg:1980ut,Schechter:1980gr,Cheng:1980qt,Mohapatra:1980yp}, e.g., the 
see-saw scale can be below the electroweak scale. The neutrino masses can 
also be generated 
radiatively~\cite{Zee:1980ai,Cheng:1980qt,Zee:1985id,Babu:1988ki,Hall:1983id,Chang:1986bp,Babu:2001ex,Babu:2002uu,Cai:2017jrq}. 
The scalars of interest to us appear in different contexts. 
For example, weak-isospin singlet 
scalars appear in radiative mass models~\cite{Zee:1985id,Babu:1988ki,Babu:2002uu}, 
whereas weak-isospin triplet scalars appear 
in light type II seesaw models and other mass models~\cite{Dev:2018sel}.
If the scalar also couples to right-handed $W^\pm$ gauge bosons, 
as in the latter case in the left-right symmetric model, 
the scalar-fermion coupling for a scalar that couples to right-handed 
electrons with a scalar mass of less than $\sim 100$ GeV 
is significantly constrained by existing 
experimental limits on neutrinoless double $\beta$ decay~\cite{Dev:2018sel}. 
This constraint does not act in our case because 
the associated scalars do not break lepton number. 
Here we suppose, as in Ref.~\cite{Gardner:2018azu}, 
 that scalars with lepton number need not 
in themselves act to explain the numerical size of
the neutrino mass, so that we take no stance on 
the precise origin of the neutrino masses and mixings. 
We consider minimal scalar models with weak-isospin triplet and singlet scalars 
that couple 
to first-generation fermions only --- such a scenario is much less constrained, 
evading severe constraints, e.g., from the $\mu$ lifetime and $\mu \to e \gamma$ 
decay~\cite{Babu:2002uu}. We do find constraints, however, on our scenario
from precision measurements of Bhabha scattering and of the $Z^0$ width.
We view minimal scalar models as a simple framework in which to study
the connections between B- and L-violating phenomena, and for scalars with masses
that would permit contributions to the $Z^0$ width we find that
it turns out to be incomplete. We also find, however, that it is simple
to remedy this and bring all into agreement 
through the addition of a higher dimensional operator, and its impact 
on the parameters of our solutions to the $\Delta a_e$ anomaly is trivially small. 
We consider these issues in Sec.~\ref{sec:other}.

We conclude this section by outlining the content of our paper --- we begin, in 
Sec.~\ref{sec:models}, by describing the scalar models we employ in more detail. 
Thereafter, in Sec.~\ref{sec:gm2e},  we discuss
the contributions to $a_e$ in these models, providing 
our detailed computations 
in the appendix for clarity. 
We describe the sets of possible couplings and masses that solve the $a_e$ anomaly
before turning to the constraints on these models
from parity-violating electron scattering in Sec.~\ref{sec:parity} 
and considering other possible constraints in Sec.~\ref{sec:other}. 
In our analysis we focus on scalars of less than ${\cal O}(10\,{\rm GeV})$ in mass, making our 
analysis complementary to that of Ref.~\cite{Dev:2018sel}, who analyzed constraints on 
doubly charged 
scalars with masses in excess of that. We conclude 
with a discussion of the experimental 
prospects in Sec.~\ref{sec:sum}. 

\section{Scalars with lepton number} 
\label{sec:models}

Minimal scalar models are extensions of the SM that respect its gauge symmetries 
and do not impact its predictive power, because the new interactions possess 
mass dimension 4 or less. 
Such models have been primarily employed in the study of baryon-number-violating 
and/or lepton-number-violating 
processes~\cite{Davies:1990sc,Bowes:1996xy,Barr:2012xb,Arnold:2012sd,Arnold:2013cva,Gardner:2018azu}, through the low-energy higher-dimension operators that can appear. 
In what follows we introduce new scalars with 
 definite representations under the 
SU(3)${}_{c}\times$SU(2)${}_{L}\times $U(1)${}_{Y}$ gauge symmetry of the SM 
that also carry nonzero lepton number L
and construct their minimal 
interactions by requiring Lorentz and SM gauge invariance. Scalars that carry
baryon number appear in this model also, and the possiblity of baryon-number-violating
proton or neutron decay is removed at tree level by choosing the particular scalars that
are allowed to appear~\cite{Arnold:2012sd,Arnold:2013cva,Gardner:2018azu}. In such
an approach the observability of the baryon-number-violating and/or lepton-number-violating
processes that can occur rest on the empirical constraints that exist on the
new scalars' masses and couplings~\cite{Gardner:2018azu}. This is in contrast to 
UV-complete models in which the gauge dynamics enforce the absence of baryon number
violation by one unit, but also admit observable neutron-antineutron oscillations.
As specific examples we note models based on the  gauge group 
${\rm SU(3)}_c \times {\rm SU(2)}_L \times {\rm SU(2)}_R \times {\rm U(1)}_{\rm {B-L}}$
\cite{Mohapatra:1980qe,Babu:2001qr}
or ${\rm SU(2)}_L \times {\rm SU(2)}_R \times {\rm SU(4)}_c$~\cite{Chacko:1998td,Babu:2008rq} 
or on the non-supersymmetric ${\rm SO(10)}$ ~\cite{Babu:2012vc}.
In these models the new light scalars 
range from about 100 GeV to the TeV scale in mass.
Thus minimal scalar models open the door to new possibilities, to the consideration 
of a relatively unexplored parameter space.
In this paper we show that new sub-GeV-scale scalars can potentially explain the
$(g-2)_e$ puzzle, but to render these solutions compatible with measurements
from LEP we need to augment our minimal scalar model with a higher dimension operator.
We refer to Sec.~\ref{sec:other} for a detailed discussion.

Generally, 
there are three possible scalars $X_i$ that couple to SM leptons only, all carrying L$=-2$. 
We have two weak isospin singlets: 
$X_1$ with hypercharge $Y=2$ that couples to 
right-handed fermions, where we employ the convention
that the electric charge $Q_{\rm em}=T_3 +Y$ in units of $|e|$ and 
$T_3$ is the third component of weak isospin, 
and $X_2$ with hypercharge $Y=1$ that couples to 
left-handed fermions. There is also one 
weak isospin triplet $X_3$ with $Y=1$ that couples to left-handed fermions. 
Since the new scalars carry electric charge to ensure
electric charge conservation, we have added scalar QED interactions
as appropriate. Through the
electroweak gauge invariant
kinetic terms, the scalars couple to the $Z^0$ gauge boson as well, and
we consider the consequences of this in Sec.~\ref{sec:other}. We will see 
that our solutions to the $(g-2)_e$ puzzle demand scalars that are lighter than
the  $Z^0$ width constraints would allow, but we find that through a small
addition to our minimal scalar models we can satisfy the $Z^0$ width 
constraint with only trivial numerical modifications to our $(g-2)_e$ solutions. 

Denoting a right-handed lepton of generation $a$ as $e^a$ 
and the associated left-handed lepton doublet as $L^a$, the possible scalar-fermion interactions
mediated by each $X_i$ are of form 
\begin{equation}
 -g_1^{ab} X_1 (e^a e^b) \,, \, 
-g_2^{ab} X_2 (L^a \varepsilon L^b) \,, \, 
-g_3^{ab} X_3^A (L^a \xi^A L^b) \,, 
\label{sff}
\end{equation}
where $\varepsilon=i\tau^2$ is a totally antisymmetric tensor, 
$\xi^A\equiv ((1+ \tau^3)/2, \tau^1 / \sqrt{2}, (1 - \tau^3)/2)$, and 
 $\tau^A$ are Pauli matrices with $A\in 1,2,3$~\cite{Gardner:2018azu}. 
The symmetries of the scalar 
representations under weak isospin SU(2) fix the 
symmetry of the associated coupling constant under $a,b$ interchange, with 
 $g_1^{ab}$ and $g_3^{ab}$ symmetric and  $g_2^{ab}$ antisymmetric. 
Thus only $X_1$ and $X_3$ can couple to first-generation leptons
exclusively. 
In Eq.~(\ref{sff}) we adopt 2-spinors such that the fermion 
products in parentheses are Lorentz invariant, and 
we map to 4-spinors via $(e_{L,R \alpha} \mu_{L,R \beta}) \rightarrow 
(e_\alpha^T C P_{L,R} \mu_\beta)$ where $C$ and 
$P_{L,R} = (1\mp \gamma_5)/2$ are in Weyl representation~\cite{Dreiner:2008tw}. 
We have chosen the arbitrary phases~\cite{Gardner:2016wov} that appear  
such that $C=i\gamma^0 \gamma^2$ and the charge-conjugate field $\psi^c$
is $\psi^c \equiv C(\overbar{\psi})^{\top}$. 
Thus the scalar-fermion interactions for each of these scalars are of form 
\begin{eqnarray}
{\cal L}_{X_1} &\supset& - g_1^{ab} X_1 \overbar{e_R^{a\,c}} e_R^b + {\rm H.c.} \,,
\nonumber\\
{\cal L}_{X_2} &\supset& - g_2^{ab} X_2 (\overbar{e_L^{a\,c}} \nu_L^b - 
\overbar{e_L^{b\,c}} \nu_L^a) + {\rm H.c.} \,,\nonumber\\
{\cal L}_{X_3} &\supset& - g_3^{ab} \left( X_3^1 \overbar{\nu_L^{a\,c}} \nu_L^b 
+ 
X_3^2 \frac{1}{\sqrt{2}}\left(\overbar{e_L^{a\,c}} \nu_L^b  + \overbar{\nu_L^{a\,c}} e_L^b \right)
+ X_3^3 \overbar{e_L^{a\,c}} e_L^b \right) + 
{\rm H.c.} \,
\label{scalf}
\end{eqnarray}
In what follows we assume that 
$X_1$ and $X_3$ couple to first-generation fermions only, whereas for $X_2$ we assume only
$1 \leftrightarrow 3$ couplings exist, since the existing constraints on 
intergenerational mixing are less severe in that case~\cite{Babu:2002uu}. 
We analyze the pertinent constraints there in Sec.~\ref{sec:other}. 

\section{New scalar contributions to $a_e$} 
\label{sec:gm2e}
\begin{figure}[tb]
\centering
\includegraphics[scale=0.45]{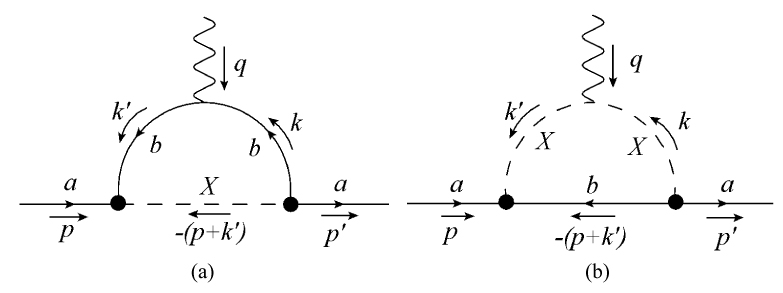}
\caption{Feynman diagrams to illustrate contributions to 
the anomalous magnetic dipole moment $a_{\ell_a}$ of lepton $a$, 
where $b$ denotes another lepton, and $X$ denotes a scalar that carries lepton number. 
Note that if lepton b is electrically neutral, only diagram (b) can contribute to 
$a_{\ell_a}$. 
} 
\label{fig:MDM}
\end{figure}

In minimal scalar models the new scalars can carry electric charge, so that 
two types of Feynman diagrams can contribute to  $a_e$ at leading order: 
one in which the photon attaches to the internal charged fermion line and a second in 
which the photon attaches to the charged scalar line --- we illustrate these possibilities
in Fig.~\ref{fig:MDM}. 
We find that $X_1$ and $X_3$ can contribute to $a_e$ through both diagrams, whereas in the
case of $X_2$ only the second diagram appears. 
The contributions to $a_e$  from $X_1$ and $X_2$ have been 
previously studied~\cite{Babu:2002uu}. 
Although we agree with Ref.~\cite{Babu:2002uu} for the computation of 
$\Delta a_e$ from $X_2$, our computation of $\Delta a_e$ from $X_1$ does not --- indeed, 
our result differs from theirs by a factor of $-4$. Consequently we find that 
the contribution to $\Delta a_e$ from each scalar is negative definite. 
Since this result is key to our paper, and subtleties exist in the computation of $\Delta a_e$, 
we present our computation in detail in the appendix.  
In this section we compile our results and evaluate their consequences. 
We evaluate the contribution of each possible new scalar to $a_e$
independently, terming this $(\delta a_e)_{X_i}$. 

Combining the results of the appendix, Eqs.~(\ref{g2a}) and (\ref{g2b}), as appropriate, 
we find that the 
 contribution to $\Delta a_e$ from $X_1$ is 
\begin{equation}
(\delta a_e)_{X_1} = 
- \frac{m_e^2 |g_1^{11}|^2}{4 \pi^2} \left( \int_0^1 dz \frac{z(1-z)^2}{ (1-z)^2 m_e^2 
+ z m_{X_1}^2}
+ 2 \int_0^1 dz \frac{z(1-z)^2}{ z^2 m_e^2 + (1-z) m_{X_1}^2} 
\right) \,,
\label{X1res}
\end{equation}
so that $(\delta a_e)_{X_1} \le 0$ and finite for all $m_{X_1} > 0$. 
Moreover, the contribution to $\Delta a_e$ from $X_2$ 
from Eq.~(\ref{g2c}) is  
\begin{equation}
(\delta a_e)_{X_2} = 
- \frac{4 m_e^2 |g_2^{1j}|^2}{16 \pi^2} \left(
\int_0^1 dz \frac{z(1-z)}{ m_{X_2}^2 -z m_e^2 } 
\right) \,,
\label{X2res}
\end{equation}
where we have set the mass of the neutrino 
$\nu_j$ to zero here and elsewhere, as it is known to be very 
small~\cite{Tanabashi:2018oca}. The $4$ in the numerator appears because 
$g_2^{1 3}=-g_2^{3 1}$, as in Ref.~\cite{Babu:2002uu}, so that there is a 2 in 
the effective $e - X_2 - \nu_j$ vertex. 
Here $M_{X_2} < m_e$ leads to a singularity 
in the parameter integral arising from on-mass-shell intermediate states; we avoid this
possibility if $M_{X_2} > m_e$. For $M_{X_2} < m_e$ we would 
replace the integral in Eq.~(\ref{X2res}) with its principal value, though 
in that region $(\delta a_e)_{X_2} > 0$.
Finally, the contribution to $\Delta a_e$ from $X_3$  is 
\begin{eqnarray}
(\delta a_e)_{X_3} &=& 
- \frac{m_e^2 |g_3^{11}|^2}{4 \pi^2} 
\Bigg( \int_0^1 dz \frac{z(1-z)^2}{ (1-z)^2 m_e^2 + z m_{X_3}^2}
+ 2 \int_0^1 dz \frac{z(1-z)^2}{ z^2 m_e^2 + (1-z) m_{X_3}^2} \nonumber \\
&&+ \frac{1}{2} \int_0^1 dz \frac{z(1-z)}{ m_{X_3}^2 -z m_e^2 } 
\Bigg) \,. 
\label{X3res}
\end{eqnarray}
Here, too, by choosing  $M_{X_3} > m_e$ 
we would avoid the inconvenience of a singularity in the parameter integral; in 
the $M_{X_3} < m_e$ region the integral would be replaced by its principal value, noting that
in this case $(\delta a_e)_{X_3} < 0$ for $M_{X_3} > 0$. 
Thus we observe that each of the three lepton-number-carrying scalars possible in minimal 
scalar models could solve the $a_e$ anomaly --- we need only choose a scalar mass and
scalar-fermion coupling consistent with the empirical value of $\Delta a_e$, and a broad range
of choices are possible. Thus we see that the $\Delta a_e$ 
anomaly could also potentially be solved by new physics at very light mass 
scales, beyond the reach of existing accelerator experiments. 
Nevertheless, in what follows we consider scalars with masses $M_{X_i} > m_e$, 
as that mass region loosely avoids astrophysical constraints, such as those from 
stellar cooling~\cite{Knapen:2017xzo}. We note, however, that 
new particles with masses $M_{X_i} < m_e$ 
may be possible if their interactions do not permit them to escape an astrophysical 
environment~\cite{Rrapaj:2015wgs} --- and our lepton-number-carrying scalars may well be of
that class. We also consider 
$M_{X_i} < 8\,{\rm GeV}$ on $X_1$ and $X_3$ because we note that existing LHC
searches for new physics in $pp$ collisions to same-sign dileptons observe
no departures from the SM but also require that the dilepton invariant mass be in excess of
$8\,{\rm GeV}$~\cite{Chatrchyan:2013fea,Khachatryan:2016kod}. 
We note that both $X_2$ and $X_3$ can induce a contribution to the 
magnetic moment of a massive  Dirac neutrino; we consider this further in Sec.~\ref{sec:other}. 

\begin{figure}[!ht]
    \centering
    \begin{subfigure}[b]{0.48\linewidth}        
        \centering
        \includegraphics[width=\linewidth]{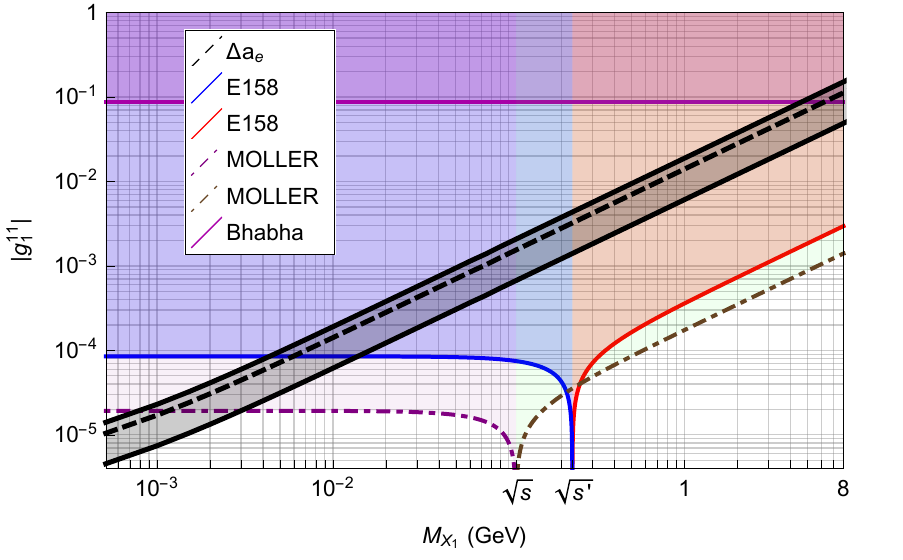}
        \caption{}
    \end{subfigure}
    \begin{subfigure}[b]{0.48\linewidth}        
        \centering
        \includegraphics[width=\linewidth]{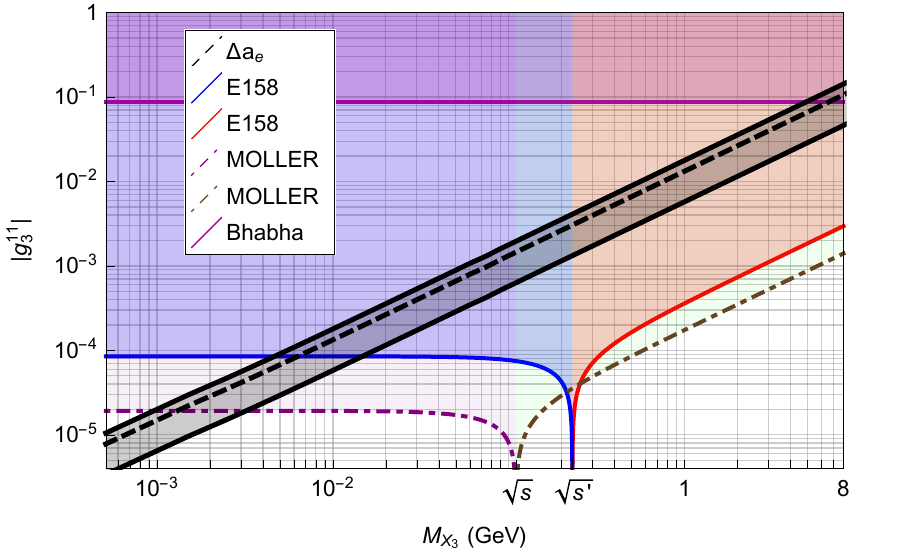}
        \caption{}
    \end{subfigure}
    \caption{Our solution for the $a_e$ anomaly in scalar mass $M_{X_i}$ 
versus the magnitude of the $X_{i} ee$ coupling, $|g_i^{11}|$, 
for scalars (a) $X_1$ and (b) $X_3$, 
compared with existing and anticipated  
experimental constraints. 
The black dashed line shows our solution 
for $\Delta a_e$ in $|g_{X_i}^{11}|$ with $M_{X_i}$ for its 
experimental central value, with the black band enclosing the solutions bounded by that
for $\Delta a_e$ taken at 95\% CL. 
Note that values of $|g_i^{11}|$ above the black band produce a 
$|\Delta a_e|$ that is too large and are thus excluded by 
the measurement; we refer to the text for further discussion. 
We also show the excluded region at 90\% CL 
from the measurement of parity-violating 
M{\o}ller scattering from the E158~\cite{Anthony:2005pm} experiment (solid boundary), 
as well as the excluded region anticipated 
from the expected sensitivity 
of the planned MOLLER experiment (dashed boundary) at Jefferson 
Laboratory~\cite{MOLLER:2008,Benesch:2014bas}, if no departure from the
SM is observed.
We also show the constraint that emerges
from measurements of Bhabha scattering at LEP~\cite{Abdallah:2005ph} evaluated
at 95\% CL --- 
see Sec.~\ref{sec:other} for a detailed discussion. }
\label{fig:soln}
\end{figure}

We now summarize our solutions for the $\Delta a_e$ anomaly. Working in the $M_{X_i} \gg m_e$ 
limit and considering $\Delta a_e$ at 95\% confidence level (CL) we find that 
the masses and scalar-fermion couplings of each $X_i$ must satisfy
\begin{eqnarray}
3.2 \times 10^{-6} \,
\le \,\, &\frac{m_e}{M_{X_1}} |g_1^{11}|&\,\, \le \, 9.7 \times 10^{-6} \,,  
\label{soln1} \\
6.5 \times 10^{-6} \,
\le \,\, &\frac{m_e}{M_{X_2}} |g_2^{1j}|& \,\,\le\, 2.0 \times 10^{-5} \,,  
\label{soln2} \\
3.4 \times 10^{-6} \,
\le \,\, &\frac{m_e}{M_{X_3}} |g_3^{11}|& \,\,\le\, 1.0 \times 10^{-5} \,,
\label{soln3}
\end{eqnarray}
where $j\ne 1$. 
We show the exact numerical solutions for $|g_i^{11}|$ and $M_{X_i}$ for $i=1,3$
in Fig.~\ref{fig:soln}, along with other pertinent constraints and their future prospects --- 
the  mass range we show is selected to evade both stellar cooling and collider bounds. 
In this mass range, $X_2$, even with the assumption of $1\leftrightarrow 3$ couplings only, 
is significantly constrained by branching ratio measurements of
semileptonic $\tau$ decay --- we update the analysis of Ref.~\cite{Babu:2002uu} in 
Sec.~\ref{sec:other}. 
We develop the established and expected constraints from parity-violating M{\o}ller scattering, 
which act on $X_1$ and $X_3$, 
in the next section. Here we wish to emphasize, in addition to providing the
solutions we have shown, that the measured value of 
$\Delta a_e$ also {\it constrains} new physics; that is, 
the upper value of Eqs.~(\ref{soln1},\ref{soln2},\ref{soln3}) 
serves as the boundary of a 95\% CL exclusion. That is, we can exclude 
\begin{eqnarray}
 &\frac{m_e}{M_{X_1}} |g_1^{11}|& \,\,>\, 9.7 \times 10^{-6} \,,  
\label{excl1} \\
 &\frac{m_e}{M_{X_2}} |g_2^{1j}|& \,\,>\, 2.0 \times 10^{-5} \,,  
\label{excl2} \\
 &\frac{m_e}{M_{X_3}} |g_3^{11}|& \,\,>\, 1.0 \times 10^{-5} \,,
\label{excl3}
\end{eqnarray}
as these regions of parameter space yield values of $|\Delta a_e|$
that are too large --- these regions, for $X_1$ 
and $X_3$, appear above the shaded black bands in 
Fig.~\ref{fig:soln}. In contrast, the regions below the
black band in Fig.~\ref{fig:soln} give values of 
$|\Delta a_e|$ that are too small --- 
although the latter region does not explain the anomaly, 
these regions of parameter space are not excluded by the $\Delta a_e$ result, 
because the scalars we have introduced need not solve 
the $\Delta a_e$ anomaly. 

\section{Constraints from parity-violating M{\o}ller scattering}
\label{sec:parity}

The parity-violating asymmetry $A_{\text{PV}}$ in 
the low-momentum-transfer scattering of longitudinally polarized electrons from 
unpolarized electrons has been measured to a precision of 
17 ppb in the E158 experiment at SLAC, yielding a determination 
of the value of 
the effective weak mixing angle $\sin^2 \theta_W^{\rm eff}$ 
to $\simeq 0.5\%$ precision~\cite{Anthony:2005pm}.  
In contrast, in 
a future experiment planned at the Jefferson Laboratory~\cite{MOLLER:2008,Benesch:2014bas}, 
the MOLLER collaboration expects to measure $A_{\text{PV}}$ to 
an overall precision of 0.7 ppb~\cite{Benesch:2014bas}, to determine 
$\sin^2 \theta_W^{\rm eff}$ to $\simeq 0.1\%$ precision~\cite{Benesch:2014bas}, 
with a commensurate improvement of $A_{\rm PV}$ as a test of new physics. 
The determination of the weak mixing angle relies 
on the theoretical assessment of $A_{\rm PV}$ 
in the SM~\cite{Derman:1979zc,Czarnecki:1995fw,Ferroglia:2003wa,Erler:2003yk,Erler:2004in,Petriello:2002wk}, 
for which electroweak radiative corrections are 
important~\cite{Czarnecki:1995fw,Ferroglia:2003wa,Erler:2003yk,Erler:2004in,Petriello:2002wk}. 
Nevertheless, as per usual 
practice~\cite{MOLLER:2008,Benesch:2014bas,Dev:2018sel}, we use the tree-level formula for 
$A_{\rm PV}$ of Ref.~\cite{Derman:1979zc} to determine the sensitivity of the 
existing and planned $A_{\rm PV}$ measurements to new physics. 
Only the doubly-charged scalars, $X_1$ and $X_3^3$, couple to two electrons, so that 
they contribute in $s$-channel to M{\o}ller scattering, i.e., via 
$e^-(p) + e^-(k)\rightarrow X_{i} \rightarrow e^-(p') + e^-(k')$. 
Since we are considering
constraints on light scalars, the value of $s$ is important, 
so that we note that 
both E158 and MOLLER are fixed-target experiments with an electron beam energy of 
$E=50$ GeV 
for E158~\cite{Anthony:2005pm} 
and $E=12$ GeV for the MOLLER experiment~\cite{Benesch:2014bas}. Thus we have 
$s\simeq 2m_eE$, with $\sqrt{s}\simeq 0.23\,{\rm GeV}$ for E158 --- we label this ``$\sqrt{s'}$''
in Fig.~\ref{fig:soln} --- and $\sqrt{s}\simeq 0.11\,{\rm GeV}$ for MOLLER. 
If a measured value of $A_{\text{PV}}$ agrees with SM
expectations, then a model-independent constraint on 
new four-electron contact interactions follows, such as those  
of either left-left or right-right form~\cite{Eichten:1983hw} 
\begin{eqnarray}
\mathcal{H}_{\rm new}  = - \frac{g^2_{\xi\xi}}{2\Lambda^2}
\overbar{\psi_\xi}\gamma^{\mu}\psi_{\xi}
\overbar{\psi_\xi}\gamma_{\mu}\psi_{\xi}
\label{new}
\end{eqnarray} 
for $\xi=L,R$. For the MOLLER experiment, e.g., 
we would have the lower bound~\cite{MOLLER:2008, Benesch:2014bas}
\begin{eqnarray}
\frac{\Lambda}{\sqrt{|g^2_{RR}-g^2_{LL}|}}=\frac{1}{\sqrt{\sqrt{2}G_F|\Delta Q^e_W|}}\simeq 7.5\ \text{TeV},\label{limit_m}
\end{eqnarray}
at 67\% CL, where $\Lambda$ is the mass scale of new physics and 
$G_F$ is the Fermi constant. We note that the error in the weak charge of the electron
$\Delta Q_W^e$, where $Q_W^e \equiv 1 - 4 \sin^2 \theta_W^{\rm eff}$~\cite{Benesch:2014bas} 
in the SM, 
 is $\pm 5.1 \times 10^{-3}$ for the E158 experiment~\cite{Anthony:2005pm} 
and is expected to be 
$\pm 1.1 \times 10^{-3}$ for the MOLLER experiment~\cite{Benesch:2014bas}. 
Interpreting both results at 90\% CL yields 
$\Lambda/{\sqrt{|g^2_{RR}-g^2_{LL}|}} \simeq 2.7\, {\rm TeV}$ and 
$\Lambda/{\sqrt{|g^2_{RR}-g^2_{LL}|}} \simeq 5.7\, {\rm TeV}$ for the E158 and 
MOLLER experiments, respectively. 

Returning to the possibility of doubly-charged scalars, we 
rewrite the interactions of Eq.~(\ref{scalf}) as 
\begin{eqnarray}
\mathcal{H} \supset 
g_i^{11} X_i\overbar{\psi^c}P_{\xi_i}\psi+g_i^{11\,*} X_i^* \overbar{\psi}P_{-\xi_i}\psi^c \,,
\end{eqnarray}
where $i$ denotes model 1 or 3. Here $\xi_1,-\xi_1$ are $R, L$  and 
$\xi_3,-\xi_3$ are $L, R$, respectively. 
Computing the $S$-matrix for M{\o}ller scattering, 
$e^-(p) + e^-(k) \rightarrow e^-(p') + e^-(k')$:
\begin{eqnarray}
\langle \pmb{p}'\pmb{k}'|T\Big(\frac{1}{2!}(-i)^2\int d^4x\ \mathcal{H}(x)\int d^4y\ \mathcal{H}(y)\Big)
|\pmb{p}\ \pmb{k}\rangle \,,
\end{eqnarray}
and noting that $\overbar{\psi^c}(x)P_{\xi}\psi(x)\overbar{\psi}(y)P_{\xi'}\psi^c(y)$ and 
$\overbar{\psi^c}(y)P_{\xi}\psi(y)\overbar{\psi}(x)P_{\xi'}\psi^c(x)$ generate the 
same contribution to the $S$-matrix, we have 
\begin{equation}
-|g_i^{11}|^2\langle \pmb{p}'\pmb{k}'|T\Big(\int d^4x\ X_i(x)\overbar{\psi^c}(x)P_{\xi}\psi(x)\int d^4y\ X_i^*(y)\overbar{\psi}(y)P_{\xi'}\psi^c(y)\Big)
|\pmb{p}\ \pmb{k}\rangle \,.
\end{equation} 
After contracting $X_i$ and $X_i^\ast$, applying a Fierz transformation~\cite{Nishi:2004st}, 
and working in the 
$s \ll M_{X_i}^2$ limit, we extract the effective Hamiltonian
\begin{eqnarray}
\mathcal{H}_{\text{eff}}=\frac{-|g_i^{11}|^2}{2M^2_i}\overbar{\psi}\gamma^{\mu}P_{\xi}\psi \overbar{\psi}\gamma_{\mu}P_{\xi}\psi\,.\label{HD}
\end{eqnarray}
Comparing with Eq.~(\ref{new}), 
we identify $g_{RR} \equiv |g_1^{11}|$ and $g_{LL} \equiv |g_3^{11}|$. For definiteness
we note that Eq.~(\ref{new}) follows from the use of the 
$Z_0$ interaction in Ref.~\cite{Derman:1979zc} to compute $A_{\rm PV}$, 
with $v=g_{RR} + g_{LL}$, $a=g_{RR} - g_{LL}$, $g_0=1/2$, which also yields 
$|g_{RR}^2 - g_{LL}^2|/\Lambda^2 \leftrightarrow \sqrt{2} G_F |\Delta Q_W^e|$ as used in 
Eq.~(\ref{limit_m}). 
Previously the relations $|g_{RR}|^2 \equiv |g_1^{11}|^2/2$ and $|g_{LL}|^2 \equiv |g_3^{11}|^2/2$
have been used to set the effective mass scale $\Lambda$ for the 
doubly-charged scalars~\cite{MOLLER:2008,Benesch:2014bas,Dev:2018sel}; however, as we
have shown, those 2's should not appear. In our current analysis we wish to constrain
light scalars, so that $s \ll M_{X_i}^2$ need no longer be satisfied. We note that 
we may still 
safely use $A_{\rm PV}$ as computed in Ref.~\cite{Derman:1979zc}
because $g_{\xi\xi}^2 s/ (2 (s -M_{X_i}^2 ) \pi\alpha) \ll 1$ can be satisfied nonetheless. 
Thus at low scales, we replace $\Lambda/\sqrt{|g^2_{RR}-g^2_{LL}|}$ by 
$\sqrt{|s-M^2_{X_i}|}/|g^{11}_{i}|$ to find the constraints
\begin{eqnarray}
\frac{\sqrt{|s-M^2_{X_i}|}}{|g^{11}_{i}|} \gtrsim 2.7\  \text{TeV}\,, \quad 
\frac{\sqrt{|s-M^2_{X_i}|}}{|g^{11}_{i}|} \gtrsim 5.7\  \text{TeV}\,,\label{Apv_limit}
\end{eqnarray}
at 90\% CL for the E158~\cite{Anthony:2005pm} and MOLLER experiments~\cite{Benesch:2014bas}, 
respectively. 
Thus we note that the constrained region depends on the center-of-mass (CM) energy for each 
experiment and that if $M_{X_i} \ll \sqrt{s}$, only the 
coupling constants $g^{11}_{i}$ are constrained. In particular, 
if $M_{X_i} \ll \sqrt{s'}$, the E158 constraint becomes 
$|g^{11}_{i}|\leq  8.58\times 10^{-5}$, whereas 
if $M_{X_i} \ll \sqrt{s}$, the MOLLER constraint becomes 
$|g^{11}_{i}|\leq  1.9\times 10^{-5}$. 
The exclusion limits from Eq.~(\ref{Apv_limit}) as a function of $M_{X_i}$ are shown 
in Fig.~\ref{fig:soln}. 
One can find that indeed 
both the solid (red) and dashed (olive) curves become straight lines as $M_{X_i}$ grows much 
bigger 
than $\sqrt{s'}$ and $\sqrt{s}$. 
Moreover, as $M_{X_i}$ becomes much smaller than $\sqrt{s'}$ and $\sqrt{s}$, 
the solid (blue) and dashed (purple) curves become flat, so that only a coupling constant
constraint emerges. (Note that the constraint from precision measurements of Bhabha scattering
at LEP~\cite{Abdallah:2005ph} is also a coupling
constant constraint because the CM energies studied
far exceed the scalar masses of interest~\cite{Swartz:1989qz}; 
we refer to Sec.~\ref{sec:other} for a detailed discussion.)
In the regions for which $M_{X_i}$ is very close to either $\sqrt{s}$ or $\sqrt{s'}$, 
the constraints of Eq.~(\ref{Apv_limit}) demand 
a very small coupling constant, though the evaluation of 
$A_{\rm PV}$ can become non-trivial --- it may be necessary to replace the scalar
propagator by a Breit-Wigner form to find a definite result. However, for the 
region shown in Fig.~\ref{fig:soln} this is not needed. 

\section{Other constraints} 
\label{sec:other} 

Light scalars that carry lepton number 
and couple to electrons, in a manner that preserves SM symmetries, also carry electric charge. 
As a result, the ``beam-dump'' experiments that severely constrain the  
electron coupling to electrically neutral, light scalars~\cite{Liu:2016qwd,Liu:2016mqv} 
do not operate, because 
electrically charged scalars interact with 
the material of the target or beam dump and do not escape. 
Certainly, too, searches 
for $s$-channel resonances in low-energy Bhabha scattering~\cite{Tsertos:1989gv} do 
not apply to the current case (and we consider the impact of new scalars in
Bhabha scattering in $t$-channel later in this section), though 
an analogous search for a low-energy, $s$-channel resonance in $e^- e^-$ scattering
should be possible, though the extremely narrow decay widths associated with the
scalar solutions we have found in Fig.~\ref{fig:soln} 
may make a sufficiently sensitive test impracticable.
In what follows we consider further 
constraints particular to scalars that carry lepton number. 

The scalars $X_1$ and $X_2$ have been previously discussed in the context of a 
particular model~\cite{Zee:1985id,Babu:1988ki} in which the neutrino 
masses are generated through radiative corrections~\cite{Zee:1980ai,Cheng:1980qt}. 
In this paper we do not delve into the origin of neutrino masses.
Nevertheless, 
the scalars $X_2$ and $X_3^2$ can potentially mediate 
additional neutrino mass contributions. We find it impossible to generate either 
a Dirac or Majorana neutrino mass at one loop level, so that our flavor-specific couplings
do not in themselves impact the neutrino mass splittings. 
However, if both $X_1$ and $X_2$ exist, then a minimal scalar interaction 
of form $\mu X_2 X_2 X_1^\ast + {\rm H.c.}$ can also exist between 
them\footnote{This interaction is the same as model F 
in our recent work~\cite{Gardner:2018azu}.}, then 
it is possble to induce a neutrino Majorana mass at two-loop 
order~\cite{Zee:1985id,Babu:2002uu, Babu:1988ki}.   
The mass prediction depends on the size of $\mu$, the 
coupling constant associated with the scalar-scalar interaction, and
although its upper bound has been estimated in Ref.~\cite{Babu:2002uu}, there are no constraints
on its minimum value --- thus these considerations do not restrict the parameter space
of interest to us in this paper. 

If neutrinos are massive Dirac particles, then 
the scalars $X_2$ and $X_3^2$ can each contribute to its
magnetic moment,
though these effects turn out to be
extremely small. 
The largest contributions in the 
region of parameter space of interest to us come from 
$X_2$ to $\mu_{\nu_\tau}$ if $M_{X_2} \simeq m_e$ and from $X_3^2$ to $\nu_e$ 
if $M_{X_3} \simeq m_e$. Employing 
Eq.~(\ref{g2c}) 
we find 
\begin{eqnarray}
[\delta\mu_{\nu_\tau}]_{X_2} 
\simeq \frac{-1}{12}\frac{|g_2^{13}|^2}{\pi^2}\frac{m^2_\nu}{m^2_{e}}\mu_B\,, \quad 
[\delta\mu_{\nu_e}]_{X_3} 
\simeq \frac{-1}{24}\frac{|g_3^{11}|^2}{\pi^2}\frac{m^2_\nu}{m^2_{e}}\mu_B\,,
\end{eqnarray}
where for simplicity we have assumed the neutrinos 
are 
approximately 
degenerate, with mass $m_\nu$ and $\mu_B$ the Bohr magneton. 
From cosmological observations, 
we have $\sum_j m_j < 0.170\,{\rm eV}$ at 95\% CL, though the best current
limit on $m_{\bar\nu_e}$ from $^{3}{\rm H}$ $\beta$-decay is 
$ m_{{\bar\nu}_e} < 2.05\,{\rm eV}$ at 95\% CL~\cite{Tanabashi:2018oca}. 
Thus we see that even with $m_\nu \simeq 2\,{\rm eV}$ and $|g_2^{13}|=1$, 
the largest contribution,
$[\delta\mu_{\nu_\tau}]_{X_2} $, 
can not be excluded by the current 
best experimental limit $|\mu|_{\nu} \sim 2.9\times 10^{-11}\mu_{B}$~\cite{Beda:2012zz},
nor by expected improvements~\cite{Giunti:2015gga,Studenikin:2016ykv}. 

\begin{figure}[!ht]
    \centering
\includegraphics[scale=0.80]{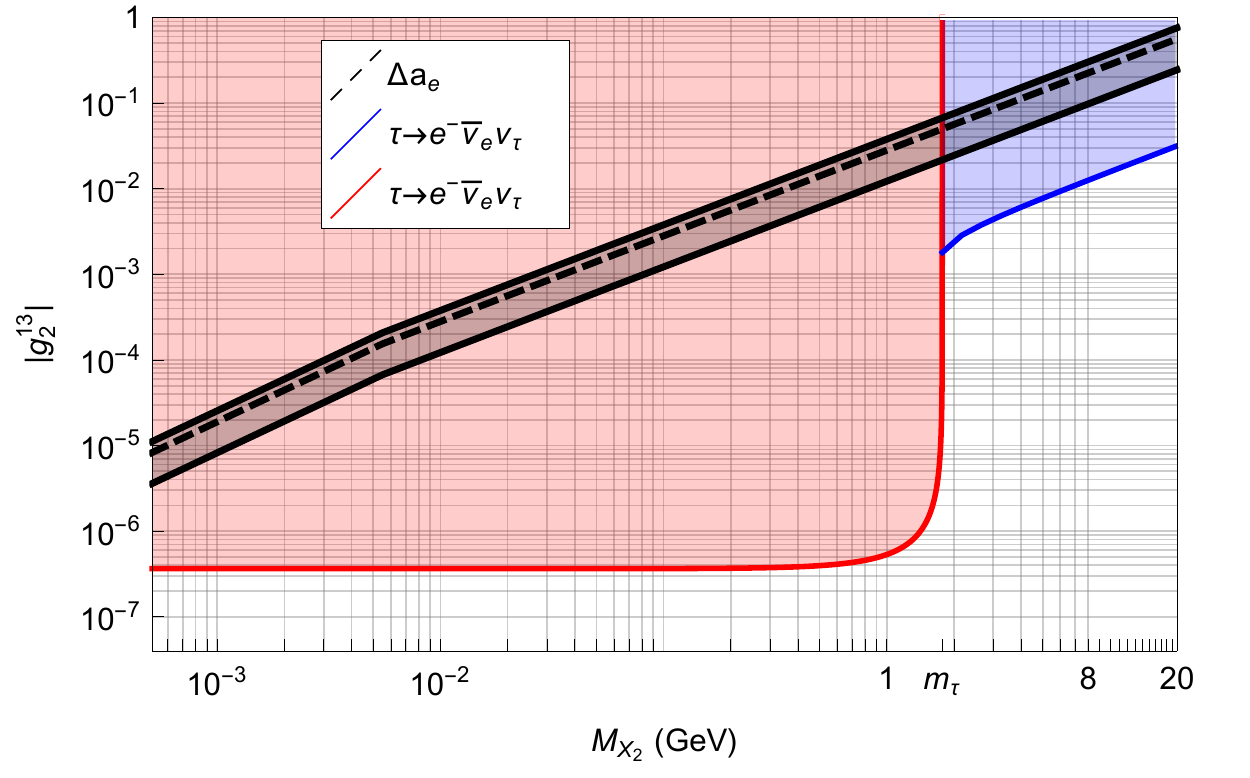}
    \caption{Our solution for the $a_e$ anomaly in scalar mass $M_{X_2}$ 
versus the magnitude of the $X_{2} e \nu_\tau$ coupling, $|g_2^{13}|$, 
compared with existing 
experimental constraints. In this case we have shown our solution over a larger 
mass range than in Fig.~(\ref{fig:soln}), because the collider constraints
on same-sign dileptons do not apply~\cite{Chatrchyan:2013fea,Khachatryan:2016kod}. 
The black dashed line and band are defined as in Fig.~\ref{fig:soln}, but 
are for $\Delta a_e$ in $|g_{2}^{13}|$ with $M_{X_2}$.
We also show the experimentally excluded region at 90\% CL from the current 
error in the measured branching 
ratio in $\tau \to e \bar\nu_e \nu_\tau$ decay~\cite{Tanabashi:2018oca}; 
for $M_{X_2} < m_\tau$ we assume 
that the $X_2$ width is saturated by $X_2 \to e^- \bar\nu_\tau$ decay 
and refer to the text for further discussion. 
}
\label{fig:soln2}
\end{figure}

We now turn to the consideration of
constraints from flavor physics, noting the comprehensive analysis of 
Ref.~\cite{ Babu:2002uu}. Taken altogether,
the constraints on flavor-non-diagonal scalar-fermion couplings 
from the experimental limits on lepton-flavor-violating processes, and from the
muon lifetime, are severe. 
As a result, we have considered first-generation couplings for $X_1$ and $X_3$, and
first-third generation couplings for $X_2$ exclusively. Consequently, we need 
only consider the constraint from the measurement of 
$\tau \to e \bar\nu_e \nu_\tau$ decay, as the only other constraint, from 
$e/\mu$ lepton-flavor universality in semileptonic $\tau$ decay, acts similarly. 

The scalar $X_2$  can mediate $\tau$ semileptonic decay via $\tau (p) \to {\bar\nu}_e X_2^\ast
\to {\bar\nu}_e (k') e^- (p') \nu_\tau (k)$. After a Fierz transformation, we find the decay
amplitude can found from the SM result by replacing 
$G_F^2 \to |g_2^{13}|^4 /[2(t - M_{X_2}^2)^2]$, where $t = (p-k')^2$. 
Working in the $\tau$ rest frame and integrating over the three-body phase space, neglecting 
all the light lepton masses, yields 
\begin{equation}
\Gamma = \frac{m_\tau |g_2^{13}|^4}{4 \pi^3}
\int_0^{m_\tau/2} d\omega' \frac{ (\omega')^2 (m_\tau - 2 \omega') }{
(m_\tau^2 - 2 m_\tau \omega' - M_{X_2}^2 )^2} \,,
\label{bigmxform}
\end{equation}
where $\omega'$ is the energy of the anti-electron neutrino. 
For $M_{X_2} > m_\tau$, the integral is well-defined, and for 
$M_{X_2} \gg m_\tau$ yields the familar result
\begin{equation}
\Gamma = \frac{m_\tau^5}{192 \pi^3} \frac{|g_2^{13}|^4}{2 M_{X_2}^4} \,.
\label{bigmx}
\end{equation}
For $M_{X_2} < m_\tau$, a $t$-channel pole appears, which we address by 
replacing the scalar propagator by a Breit-Wigner form: 
\begin{equation}
\frac{1}{(t - M_{X_2})^2} \to \frac{1}{| t - M_{X_2}^2 + i M_{X_2} \Gamma_{X_2} |^2} \,.
\label{scalartobw}
\end{equation}
Defining 
\begin{equation}
x = \frac{t}{m_\tau^2} \quad , \quad x_X = \frac{M_{X_2}^2}{m_\tau^2} \quad , \quad 
\tilde{\Gamma}_X = \frac{\Gamma_{X_2}}{m_\tau} \,,
\end{equation}
we thus have 
\begin{equation}
\Gamma = \frac{m_\tau |g_2^{13}|^4}{32 \pi^3}
\int_0^{1} dx \frac{ (1-x)^2 x}{  (x-x_X)^2 + x_X \tilde \Gamma_X^2 } \,.
\end{equation}
Since $x_X \tilde \Gamma_X^2 \ll 1$, we can apply the narrow 
width approximation~\cite{Gardner:2015wea}, i.e., 
\begin{eqnarray}
\Big((x-x_{A'})^2+x_{A'}\tilde{\Gamma}^2_{A'}\Big)^{-1} \rightarrow \frac{\pi}{\sqrt{x_{A'}}\tilde{\Gamma}_{A'}}\delta(x-x_{A'}),
\end{eqnarray}
to find 
\begin{equation}
\Gamma = \frac{m_\tau |g_2^{13}|^4}{32 \pi^2} \frac{m_\tau M_{X_2}}{\Gamma_{X_2}}
\left( 1 - \frac{M_{X_2}^2}{m_\tau^2} \right)^2 \,.
\end{equation}
Since there is only one decay channel left for 
$X_2$, $X_2^\ast \to e^- {\nu}_{\tau}$, we compute 
\begin{eqnarray}
\Gamma_{X_2}  = \frac{1}{4\pi} M_{X_2} |g_2^{13}|^2 
\end{eqnarray}
to find 
\begin{eqnarray}
\Gamma = \frac{m_\tau |g_2^{13}|^2}{8 \pi} 
\left( 1 - \frac{M_{X_2}^2}{m_\tau^2} \right)^2 \,, 
\label{smamx}
\end{eqnarray}
which, as expected, is identical to our result for 
$\Gamma(\tau \to e X_2^\ast)$. We now turn to the numerical constraints on the
scalar-fermion couplings with $M_{X_2}$, given existing measurements of the 
$\tau \to e \bar \nu_e \nu_\tau$ branching ratio and $\tau$ lifetime. 
Referring to Ref.~\cite{Tanabashi:2018oca} for all 
experimental 
parameters, we note particularly that 
${\rm Br}(\tau \to e \bar\nu_e \nu_\tau) = 17.82 \pm 0.04$~\%, 
$\tau_\tau = (290.3 \pm 0.5)\times 10^{-15}\,{\rm s}$, and 
$m_\tau = 1776.86 \pm 0.12 \,{\rm MeV}$. For $M_X \gg m_\tau$, we 
can constrain, at 90\% CL, 
\begin{equation}
\frac{|g_2^{13}|^4}{2 M_{X_2}^4 G_F^2} \le 
\frac{\eta}{{\rm Br}(\tau \to e \bar\nu_e \nu_\tau)} 
\Longrightarrow 
\frac{|g_2^{13}|}{M_{X_2}} \le 1.0 \times 10^{-3}\, {\rm GeV}^{-1}
\end{equation}
or 
\begin{equation}
\frac{m_\tau^5 }{192 \pi^3} \frac{|g_2^{13}|^4}{2 M_{X_2}^4 }
\le \frac{\eta h}{100 \tau_\tau} 
\Longrightarrow 
\frac{|g_2^{13}|}{M_{X_2}} \le 1.6 \times 10^{-3}\, {\rm GeV}^{-1} \,, 
\label{thisone}
\end{equation}
with $\eta=0.066$. 
The two estimates differ in that the former implicitly assumes the leading-order formula 
describes the SM decay rate, though various refinements exist~\cite{Pich:2013lsa}. 
We note that the numerical limit reported by Ref.~\cite{Babu:2002uu} in this case is significantly 
more severe than what we report. 
In what follows we use our second method to determine the exclusion limit. 
For $M_{X_2} > m_\tau$ we replace the left-hand side (LHS) of Eq.~(\ref{thisone})
with Eq.~(\ref{bigmxform}). For $M_{X_2} < m_\tau$ 
we replace the LHS of Eq.~(\ref{thisone}) with Eq.~(\ref{smamx}). We report the 90\% CL 
exclusion we 
have found in Fig.~\ref{fig:soln2}, recalling that $(\delta a_e)_{X_2} < 0$ only if 
$M_{X_2} > m_e$. Thus we see that in this case the existing empirical data rules out
$X_2$ as a solution to the $a_e$ anomaly, at least in a minimal scalar model. 
More generally, we note that Eq.~(\ref{smamx}) can be written~\cite{Gardner:2015wea}
\begin{equation}
\Gamma = \frac{m_\tau |g_2^{13}|^4}{8 \pi} 
\left( 1 - \frac{M_{X_2}^2}{m_\tau^2} \right)^2 {\rm Br}(X_2^\ast \to e^- \nu_\tau)\,
\end{equation}
and that decreasing ${\rm Br}(X_2^\ast \to e^- \nu_\tau)$ from unity weakens
the constraint on $|g_2^{13}|/M_{X_2}$ in the $M_{X_2} < m_\tau$ region. 

Finally, we turn to the constraints that appear because our scalars  
couple to the gauge bosons of the SM.
The doubly-charged scalars that
we consider are constrained just as 
the doubly charged Higgs bosons
${H}_{L, R}^{\pm\pm}$~\cite{Mohapatra:1980yp,GELMINI1981411,Mohapatra:1981pm,Barger:1982cy}
in generalized left-right symmetric models~\cite{Grifols:1989xe} 
are. In what follows, the constraints on $H_R^{\pm\pm}$ ($H_L^{\pm\pm}$) 
are identical to those on
$X_1$ ($X_3^3$). 
We note that the same-sign dilepton
limits from searches for $pp\, [q{\bar q}]\to H_{L,R}^{\pm \pm} H_{L,R}^{\mp \mp}
\to \ell^\pm \ell^\pm \ell^\mp \ell^\mp$
from the LHC at $\sqrt{s} =13\,{\rm TeV}$ with $\ell \in e,\mu$
yield $M_{{H}_L^{++}} > 768\,{\rm GeV}$ at 95\% CL and
$M_{{H}_R^{++}} > 658\,{\rm GeV}$ at 95\% CL for
${\rm Br}({H}^{++}_{L,R} \to e^+e^+)=1$~\cite{Aaboud:2017qph,CMS},
where the experiments are most sensitive to doubly-charged scalars
with masses in excess of $200\,{\rm GeV}$ --- 
e.g., the same-sign dilepton invariant mass
is required to be in excess of $200\,{\rm GeV}$ in the study of the
$e^\pm e^\pm e^\mp e^\mp$ final state~\cite{Aaboud:2017qph}.
Thus to
constrain lighter mass scalars we must look further. 
Extensive searches for charged scalars have been made at
LEP~\cite{Abbiendi:2013hk}.  
Such measurements can probe doubly charged scalars over a very wide mass range,
both indirectly, through $t$-channel exchange of $H^{\pm\pm}$ in
Bhabha scattering scattering~\cite{Swartz:1989qz,Abbiendi:2003pr,Abdallah:2005ph},
and directly, through associated production, 
$e^+e^- \to e^\pm e^\pm {H}^{\mp\mp}$~\cite{Abbiendi:2003pr}. The
latter process tends to be more sensitive to the size of the Higgs coupling to electrons
$h_{ee}$ (our $g_1^{11}$ or $g_3^{11}$),
but the former is sensitive to a much broader range of masses.
In these experiments no evidence for the existence of ${H}^{\pm\pm}$ has been found,
with an upper limit of 
$h_{ee}<0.071$ at 95\% CL inferred for $M_{H^{\pm\pm}} < 160\,{\rm GeV} $ coming from
their direct search, 
though the region with $M_{H^{\pm\pm}} < 98.5\,{\rm GeV}$ had been
assumed to be excluded by studies of pair production. In particular, the direct 
search did not search for doubly charged scalars less than 80 GeV in
mass~\cite{Abbiendi:2003pr}. 
Turning to the pair production studies, 
through $e^+e^-$ scattering in $s$-channel~\cite{Abbiendi:2001cr}, 
a mass limit of $98.5\,{\rm GeV}$ at 95\% CL has indeed been set, but 
a lower mass limit of $45\,{\rm GeV}$ is  assumed from $Z^0$ decay
studies~\cite{Abbiendi:2001cr}. We note that 
doubly charged scalars have been studied in $Z^0$ decay, 
$Z^0 \to {H}^{++} {H}^{--}$~\cite{Acton:1992zp}.
The experiment is unable to constrain scalars with masses of 
less than a few GeV directly, and constraints on the mass of 
${H}^{\pm\pm}$ are found by appealing to measurements
of the $Z^0$ line shape. That is, they determine that the difference between
the $Z^0$ width measurement and its SM prediction to be less than 40 MeV at
95\% CL, so that a bound on the doubly charged scalar mass is set by requiring
that the $Z^0 \to {H}^{++} {H}^{--}$ partial width to be no larger than
$40\, {\rm MeV}$~\cite{Acton:1992zp}.
In this way they finally determine the mass exclusion
limits of less than 25.5 GeV for weak-isospin singlets (our $X_1$) and of
less than 30.4 GeV for weak-isospin triplets (our $X_3$) at 95\% CL~\cite{Acton:1992zp}
using~\cite{Grifols:1989xe}
\begin{equation}
  \label{ZtoHH}
  \Gamma(Z^0 \to H^{++} H^{--})
  =\frac{G_F M_Z^3}{6\pi\sqrt{2}} \left(I_3^L - Q \sin^2\theta_W\right)^2
  \left(1 - \frac{4 M_H^2}{M_Z^2}\right)^{3/2} \,,
\end{equation}
where $M_H$, $Q$, $I_3^L$ are the mass, charge, and the third component of weak isospin of the
$H^{\pm\pm}$. For the right-handed singlet we set $I_3^L=0$. We can easily
mitigate this constraint, however, through an addition to our model, as we detail below. 
There is also a
pair production constraint
extracted from 
$e^+ e^- \to e^+ e^- \ell^+ \ell^-$ data
measured by the CELLO collaboration at PETRA to realize tests of QED~\cite{LeDiberder:1988eu}, 
which Swartz has analyzed to determine a 
limit of 21.5 GeV at 90\% CL on the mass of the doubly-charged scalar if 
${\rm Br}(H^{\pm\pm} \to e^\pm e^\pm)=1$~\cite{Swartz:1989qz}. The decay width
of the doubly charged scalar is given by~\cite{Swartz:1989qz}
\begin{equation}
  \Gamma_{\ell \ell} = \frac{h_{\ell \ell}^2 }{8\pi} M_H
  \left(1 - 2\frac{m_\ell^2}{M_H^2}  \right) \left(1 - 4\frac{m_\ell^2}{M_H^2}  \right)^{1/2} \,,
\end{equation}
and Ref.~\cite{Swartz:1989qz} notes that the doubly charged scalar can be short-lived
unless $h_{\ell \ell} < 10^{-9}$. However, this observation does not bear out for
lighter mass scalars. In the empirical study of $e^+ e^- \to e^+ e^- \ell^+ \ell^-$
by Le Diberder~\cite{LeDiberder:1988eu}, three out of the four final state leptons
were detected under the requirement of a ``good vertex'' (as per Eq.~(A-1.3) of
Ref.~\cite{LeDiberder:1988eu}) in order to control backgrounds.
As a result a produced doubly charged scalar with a decay length in excess of
$0.4\,{\rm cm}$ would not have been detected by the experiment. We find that this
requirement removes light, weakly coupled scalars from the aforementioned
constraint. Namely, requiring that
 the decay length in the laboratory
 frame satisfies 
\begin{equation}
  \left(  \frac{\sqrt{s}}{2M_H} \sqrt{1 - \frac{4 M_H^2}{s}} \right)
  \frac{\hbar c}{\Gamma_{\ell \ell}} <
  0.4\, {\rm cm}\,,
\end{equation}
we see that for $\sqrt{s}$ of $40\,{\rm GeV}$\footnote{The experiment employed beam energies
from $17.5$ to $23\, {\rm GeV}$~\cite{Swartz:1989qz}.}, e.g., 
if $M_H = 1\,{\rm  GeV}$ then couplings with $h_{ee} >  5.0 \cdot 10^{-6}$ are excluded, 
whereas if $M_H = 100\, {\rm  MeV}\, (10\, {\rm MeV})$ then the exclusion limit changes to
  $h_{ee} > 5.0 \cdot 10^{-5} (5.0 \cdot 10^{-4})$. 
Thus we observe that our possible $(g-2)_e$ solutions are not constrained by the PETRA data.
From our discussion we observe that the only significant constraint on the mass
of the light scalar comes from the  measured width of the $Z^0$ gauge boson. 

Further constraints come from the indirect process, Bhabha scattering. In this
case, if $M_H^2 \ll s$~\cite{Swartz:1989qz}, 
the indirect process becomes insensitive to the mass of the scalar, 
much as we have seen in the case of M{\o}ller scattering, constraining only 
the $h_{ee}$ coupling constant in this limit. We note the limit of
$h_{ee} < 0.14$ at 95\% CL
from  
$e^+e^- \to e^+ e^- $ collision data at 
CM energies of $\sqrt{s} = 183-209\,{\rm GeV}$
collected by the OPAL detector~\cite{Abbiendi:2003pr}. 
Moreover, $e^+ e^- \to e^+ e^-$ cross section measurements 
at CM energies of $\sqrt{s} \sim 130-207\,{\rm GeV}$ at LEP by the 
DELPHI collaboration yield a limit of $h_{ee} < 0.088$ at 95\% CL,
determined from their
limit on a new contact interaction of the form in Eq.~(\ref{HD}), 
with $M_i \equiv \Lambda^-=6.8\,{\rm TeV}$ for LL and RR from Table 30
for a coupling of strength $g=\sqrt{4\pi}$~\cite{Abdallah:2005ph}.
This last limit is reported in Fig.~\ref{fig:soln}. 

Finally, since our scalars can carry electric charge, 
we evaluate the indirect constraints on them that follow from the direct measurement of the 
running of $\alpha (s)$, $|\alpha(s)/\alpha(0)|^2$, where $\alpha\equiv \alpha(0)$. 
This can be determined 
from the measured differential cross section for $e^+e^- \to \mu^+ \mu^- \gamma$, 
for which the 
most precise results are in the time-like region below 1 GeV~\cite{KLOE-2:2016mgi} --- there 
the presence of hadronic contributions is established at more than 5$\sigma$. 
We evaluate 
$\alpha(s) = \alpha/(1 - \Delta\alpha (s))$~\cite{Jegerlehner:2001wq}, 
where the leading contribution to the vacuum polarizaton $\Delta\alpha$  can 
be readily calculated in scalar QED to yield~\cite{Peskin:1995ev} 
\begin{eqnarray} 
{\rm Re}\Delta\alpha_{_{X_2}} (s) &=& - \frac{\alpha}{2\pi} \int_0^1 dx\, x(2x-1) 
\log \Big| \frac{M_{X_2}^2}{M_{X_2}^2 -s x (1-x)} \Big| \,,\nonumber \\
&=& 
 \frac{\alpha}{12\pi} \Bigg[ \log \Big( \frac{s}{M_{X_2}^2} \Big) - \frac{8}{3} \Bigg] \quad 
\text{for} \,  s \gg 4 M_{X_2}^2 \,; \\
{\rm Im}\Delta\alpha_{_{X_2}} (s) &=& 
- i \frac{\alpha}{12} \Big( 1 - \frac{4M_{X_2}^2}{s} 
\Big)^{3/2} \Theta(s - 4 M_{X_2}^2) \,.  
\end{eqnarray} 
We note that $\Delta\alpha_{_{X_2}} (s)$ is 4 times smaller,  and 
${\rm Re} \Delta\alpha_{_{X_2}} (s)$ 
runs more slowly, than that for a fermion in QED. 
The contribution of $X_2$ for $M_{X_2} \le m_e$ to 
$|\alpha(s)/\alpha(0)|^2$ 
deviates from unity by less than 0.5\% over the 
$s$-range of the experiment, $0.6 < \sqrt{s} < 0.975\,{\rm GeV}$, with an 
inappreciable $s$ dependence. Since the individual measurements 
have a statistical error of $\le 1 \%$ and an overall 
systematic error of 1\%~\cite{KLOE-2:2016mgi}, 
the existence of the $X_2$ scalar is not constrained. 
However, the contributions from $X_1$ and $X_3$ include
doubly charged scalars, and we have 
$\Delta\alpha_{_{X_1}} (s) = 4 \Delta\alpha_{_{X_2}} (s) $ and 
$\Delta\alpha_{_{X_3}} (s) = 5 \Delta\alpha_{_{X_2}} (s) $. 
Although the contributions to $\alpha(s)$ from $X_1$ and $X_3$ also have negligibly small 
slope in the $s$-range of interest, they can each generate an appreciable
offset from zero. We suppose that the existence of these scalars is limited
by the size of the overall systematic error, or offset, in the measurement of 
$|\alpha(s)/\alpha(0)|^2$. Noting the measured data points and their errors in 
Table 2 of Ref.~\cite{KLOE-2:2016mgi}, we require that the overall shift in the
theory contribution with a new scalar 
be less than $0.011$ for $\sqrt{s} < 0.783\,{\rm GeV}$, the region for 
which the hadronic contribution is completely captured by the 
included $2\pi$ intermediate state. 
Thus we estimate 
$M_{X_1} > 8.4\,{\rm  MeV}$ and 
$M_{X_3} > 19\,{\rm MeV}$. 
We regard these limits as guidelines rather than 
exclusions because the new scalars generate 
contributions that do not impact the measured $s$ dependence, 
but, rather, only its overall normalization. 
Nevertheless, this analysis suggests that $X_1$ is a more likely solution to the
$(g-2)_e$ anomaly. 

We have found severe constraints on the allowed doubly-charged scalar mass from its
couplings to
the $Z^0$ and to the photon, notably through the running of $\alpha$.
We note that the $Z^0$ constraint, in particular,
can be readily mitigated through the introduction of
a higher dimension operator that acts to neutralize
the couplings of the doubly charged scalars to SM gauge bosons. That is, we can 
add an operator of form
\begin{equation}
  - g_\Phi \frac{|\Phi|^2 |D_\mu X_i|^2}{\Lambda_\Phi^2} \,,
  \label{neut} 
\end{equation}
where the scalar $\Phi$ is an electroweak singlet with zero L and zero electric charge.
We let $\Phi$ gain a vacuum expectation value $v_\Phi$ below the scale
$\Lambda_\Phi$, where $v_\Phi \sim \Lambda_\Phi$ exceeds the electroweak scale, 
the coefficient $-g_\Phi v_\Phi^2/\Lambda_\Phi^2$, with $g_{\Phi} >0$,   acts to neutralize
the lepton-number-carrying scalars' SU(2)$_L$ and electric charges. 
Turning to Eq.~(\ref{ZtoHH}) and considering the limit on $X_1$, under which, e.g., 
the factor $(1 - 4 M_H^2 / M_Z^2)^{3/2}$ evaluates to 0.58, we see that by weakening 
the effective SU(2) coupling of the $X_1$
by about 20\% we would be
able to remove this constraint completely. This seems plausibly attainable, and
we note that such a change makes  only a trivially small impact on the
$\Delta a_e$ solutions we show in Fig.~\ref{fig:soln} because the contributions of
the charged scalars themselves to $a_e$ are numerically quite small. Thus
we have not included this effect in Fig.~\ref{fig:soln}. 

\section{Summary} 
\label{sec:sum}

In this paper we have shown that the light scalars with lepton number that appear in 
minimal scalar models of new physics can generate solutions to the $\Delta a_e$ 
anomaly, in that they act to reduce the size of $|a_e|$. Although our solutions 
determine only the ratio of the scalar-fermion coupling to mass, we have particularly focussed 
on new particles with masses in excess of the electron mass and less than 8 GeV, 
as this mass region, at first glance, 
should evade both astrophysical cooling constraints and collider bounds. 
We should note, however, that since 
the scalars that couple to electrons also carry electric charge, 
lighter mass candidates could also prove phenomenologically viable, because such 
particles may be unable to escape an astrophysical environment and contribute to its cooling.
We have proposed three possible solutions to the $\Delta a_e$ anomaly, but we have
found that only the two solutions with doubly charged scalars are viable, because the 
existing $\tau$ decay data preclude the singly-charged scalar $X_2$ 
as a possible solution, at least in a minimal scalar model. 
As for the doubly-charged scalars, the constraints 
from parity-violating M{\o}ller scattering permit a solution to the $\Delta a_e$
anomaly, with the upcoming MOLLER experiment poised to discover a conflict with 
the SM or to constrain our proposed solutions yet further. We have also
carefully studied existing collider constraints on doubly charged scalars,
and we have noted that the only pertinent constraint on  the solutions we
consider comes from studies of the $Z^0$ line width. We can readily weaken this
constraint as needed 
through the addition of a higher dimension operator that acts to neutralize
the SU(2) and electric charges of the doubly charged scalar boson, and this 
addition leaves the parameters of our proposed $\Delta a_e$ solutions
essentially unchanged.

We have noted, moreover, that the $\Delta a_e$ determination also 
constrains broad swatches of the scalar-fermion coupling and mass 
parameter space, as parameters which would give too large a value of 
$|a_e|$ should be excluded. There are plans to make substantially 
improved measurements of both the electron and the positron anomalous
magnetic moments~\cite{Gabrielse:2019cgf}, to better existing measurements
by a factor of 10 and 150~\cite{dehmelt1987}, respectively. 
Although this comparison is meant as a CPT test, it can also help 
affirm our new physics solution to the $\Delta a_e$ anomaly, as 
the two new measurements could well agree with each other, up to the
expected difference in overall sign, but yet disagree with the SM using $\alpha$
determined through atom interferometry. The scalar solutions we have found 
can also help engender baryon and lepton number violation in 
low-energy scattering experiments, and we keenly await these studies. 
 
\ssection{Acknowledgments}
We acknowledge partial support from the Department
of Energy Office of Nuclear Physics under 
contract DE-FG02-96ER40989. We thank Mark Pitt for alerting us to the importance
of parity-violating electron scattering in this context. 
We also thank Yu-Sheng Liu for his generous assistance with the exclusion plots, 
and Heather Logan and Daniel Stolarski for prompting us to investigate the 
constraints from the running of $\alpha$ carefully.
We thank Brian Batell, Bhupal Dev, and Tao Han for key input regarding the
constraints from LEP on our model, and we thank 
Jeffrey Berryman and Yue Zhang for helpful discussions as well. 

\setcounter{equation}{0}
\renewcommand\theequation{A.\arabic{equation}}
\section*{Appendix}
\label{sec:appA}

Herewith we detail our $a_\ell$ 
computation for scalars that carry lepton number. 
The nature of the scalar-fermion interactions in this case, Eq.~(\ref{scalf}), allows 
for multiple ways in which the fermion fields can contract, so that it is more 
efficient to evaluate the time-ordered products of fields directly, rather
than to develop Feynman rules for this case. 

We have defined $\psi^c$ as $\psi^c \equiv C(\overbar{\psi})^{\top}$, noting 
the charge conjugation matrix $C$ obeys 
\begin{equation}
C^{\top}=C^{\dagger}=C^{-1}=-C\,,\label{matrix_C}
\end{equation}
as well as 
\begin{eqnarray}
C(\gamma^{\mu})^{\top}=-\gamma^{\mu}C, \quad C(\sigma^{\mu\nu})^{\top}=-\sigma^{\mu\nu}C\,.\label{C_prop}
\end{eqnarray}
We first summarize the plane-wave expansions of a Dirac field $\psi(x)$, its charge conjugate
$\psi^c(x)$, and their Dirac adjoints, 
where we refer to Ref.~\cite{Peskin:1995ev} for all details: 
\begin{eqnarray}
\psi(x)&=&\int\frac{d^3\pmb{p}}{(2\pi)^3}\frac{1}{\sqrt{2E}}\sum_{s}
\Big(a^s_{\pmb{p}}u(s,p)e^{-ip\cdot x}+b^{s\dagger}_{\pmb{p}}v(s,p)e^{ip\cdot x}\Big)\,, \\
\overbar{\psi}(x)&=&\int\frac{d^3\pmb{p}}{(2\pi)^3}\frac{1}{\sqrt{2E}}\sum_{s}
\Big(a^{s\dagger}_{\pmb{p}}\overbar{u}(s,p)e^{ip\cdot x}+b^{s}_{\pmb{p}}\overbar{v}(s,p)e^{-ip\cdot x}\Big)\,, \\
\psi^c(x)&=&\int\frac{d^3\pmb{p}}{(2\pi)^3}\frac{1}{\sqrt{2E}}\sum_{s}
\Big(a^{s\dagger}_{\pmb{p}}u^c(s,p)e^{ip\cdot x}+b^{s}_{\pmb{p}}v^c(s,p)e^{-ip\cdot x}\Big)\,, \\
\overbar{\psi^c}(x)&=&\int\frac{d^3\pmb{p}}{(2\pi)^3}\frac{1}{\sqrt{2E}}\sum_{s}
\Big(a^s_{\pmb{p}}\overbar{u^c}(s,p)e^{-ip\cdot x}+b^{s\dagger}_{\pmb{p}}\overbar{v^c}(s,p)e^{ip\cdot x}\Big)\,. 
\end{eqnarray}
We note $u^c$ and $v^c$ are defined in the manner of $\psi^c$, 
and the creation and annihilation operators obey the anticommutation relations
\begin{equation}
\{a^r_{\pmb{p}}, a^{s\dagger}_{\pmb{q}}\}=\{b^r_{\pmb{p}}, b^{s\dagger}_{\pmb{q}}\}=
(2\pi)^3\delta^3(\pmb{p}-\pmb{q})\delta^{rs}\,.
\end{equation}
We now summarize all the Wick contractions that can appear. 
The contractions 
between $\psi(x)$, $\overbar{\psi}(x)$, $\psi^c(x)$, and $\overbar{\psi^c}(x)$ and 
an incoming or outgoing fermion of mass $m$ are
\begin{eqnarray}
  \contraction{}{\psi}{{}_a(x)|\pmb{p}}{,} \psi_a(x) |\pmb{p},s \rangle = u_a(s,p) e^{-ip\cdot x}|0\rangle &\quad& \langle 
 \pmb{p} \contraction{}{\vphantom{\overbar\psi},}{s|}{\overbar{\psi}} ,s|\overbar{\psi}_a(x) 
  = \langle 0|\overbar{u}_a(s,p)e^{ip\cdot x}\,, \\  
  \contraction{}{\overbar{\psi^c_a}}{(x)|\pmb{p}}{,} \overbar{\psi^c_a}(x)|\pmb{p},s \rangle = \overbar{u^c_a} (s,p)e^{-ip\cdot x}|0\rangle 
  &\quad& \langle 
 \pmb{p} \contraction{}{\vphantom{\psi^c},}{s|}{\psi^c} ,s|\psi^c_a(x) 
  = \langle 0|u^c_a(s,p)e^{ip\cdot x}\,, 
\end{eqnarray}
where 
$|\pmb{p},s\rangle=\sqrt{2E_{\pmb{p}}}a^{s\dagger}_{\pmb{p}}|0\rangle$ and $\langle\pmb{p},s|$ denote an incoming 
and an outgoing fermion with momentum $\pmb{p}$ and spin $s$, respectively, whereas
the contractions to an incoming or outgoing antifermion are 
\begin{eqnarray}
  \contraction{}{\overbar{\psi}}{{}_a(x)|\pmb{k}}{,} \overbar{\psi}_a(x)|\pmb{k},r \rangle = \overbar{v}_a (r,k)e^{-ik\cdot x}|0\rangle 
  &\quad& \langle 
  \pmb{k} \contraction{}{\vphantom{\psi},}{r|}{\psi} ,r|\psi_a(x) 
  = \langle 0|v_a(r,p)e^{ik\cdot x}\,, \\ 
   \contraction{}{\psi}{{}^c_a(x)|\pmb{k}}{,} \psi^c_a(x) |\pmb{k},r \rangle = v^c_a(r,k) e^{-ik\cdot x}|0\rangle &\quad& \langle 
 \pmb{k} \contraction{}{\vphantom{\overbar{\psi^c}},}{r|}{\overbar{\psi}} ,r|\overbar{\psi^c_a}(x) 
  = \langle 0|\overbar{v^c_a}(r,k)e^{ik\cdot x}\,, 
\end{eqnarray}
where, similarly, $|\pmb{k},r\rangle$ ($\langle\pmb{k},r|$) denote an incoming (outgoing) 
anti-fermion with momentum $\pmb{k}$ and spin $r$. 
The spinor index $a$ runs from 1 to 4. 
Different contractions of the internal fermion and antifermion fields can appear. 
That is, 
\begin{eqnarray}
&&\contraction{}{\psi}{{}_a(x)}{\overbar{\psi}} \psi_a(x)\overbar{\psi}_b(y)=
\int\frac{d^4p}{(2\pi)^4}\frac{i(\slashed{p}+m)_{ab}}{p^2-m^2+i\epsilon}e^{-ip\cdot(x-y)}\,, \\
&&\contraction{}{\psi}{{}_a(x)}{\psi} \psi_a(x)\psi^c_b(y)=
\int\frac{d^4p}{(2\pi)^4}\frac{i[(\slashed{p}+m)C^{\top}]_{ab}}{p^2-m^2+i\epsilon}e^{-ip\cdot(x-y)}\,, \\ 
&&\contraction{}{\overbar{\psi^c_a}}{(x)}{\overbar{\psi}}\overbar{\psi^c_a}(x)\overbar{\psi}_b(y)=
\int\frac{d^4p}{(2\pi)^4}\frac{i[C^{\top}(\slashed{p}+m)]_{ab}}{p^2-m^2+i\epsilon}e^{-ip\cdot(x-y)}\,, \\ 
&&\contraction{}{\overbar{\psi^c_a}}{(x)}{\psi}\overbar{\psi^c_a}(x)\psi^c_b(y)=
\int\frac{d^4p}{(2\pi)^4}\frac{i[C^{\top}(\slashed{p}+m)C^{\top}]_{ab}}{p^2-m^2+i\epsilon}e^{-ip\cdot(x-y)}\,,
\end{eqnarray}
where $a$ and $b$ are spinor indices.  

We now can compute the one-loop amplitude associated with the lepton anomalous magnetic dipole 
moment $a_\ell$. As shown in Fig.~\ref{fig:MDM}, 
a photon can be attached to either a charged fermion line or a charged scalar line, and
interactions from quantum electrodynamics (QED) and scalar QED are needed: 
\begin{eqnarray}
&& \mathcal{H}_{1}\supset - Qe\overbar{\psi}\gamma^{\mu}\psi A_{\mu} \,, \label{QED} \\ 
&& \mathcal{H}_2\supset - iQe[(\partial^{\mu}X)X^{*}-X(\partial^{\mu}X^*)]A_{\mu} \,, \label{SQED}
\end{eqnarray} 
where $Q=-1$ for the electron. Noting Eq.~(\ref{scalf}), we make the replacements 
$g_i^{11} \to g_i$ for $i=1,3$ and $e \to \psi$. Here we consider the 
contributions from $X_1$ and $X_3^3$. We address the 
contribution to $a_\ell$ from $X_2$, as well as from $X_3^2$, later. 

For the first case, the interaction is 
\begin{eqnarray}
\mathcal{H} \supset  - eQ\overbar{\psi}\gamma^{\mu}\psi A_{\mu}+
g_iX_i\overbar{\psi^c}P_{\xi}\psi+g_i^* X_i^* \overbar{\psi}P_{\xi'}\psi^c \,,
\end{eqnarray}
where 
$P_{\xi}=(1+\xi\gamma^5)/2$ is the chiral projection operator with 
$\xi=\pm1$ for R or L. 
Hermitian conjugation of the second term results in the third term, in which 
$\xi'=-\xi$.   
The one-loop contribution 
comes from the $H^3$ term of the $S$-matrix: 
\begin{eqnarray}
\langle \pmb{p}'|T\Big(\frac{1}{3!}(-i)^3\int d^4x\ \mathcal{H}(x)\int d^4y\ \mathcal{H}(y)\int d^4z\ \mathcal{H}(z)\Big)
|\pmb{p}\ \pmb{q}\rangle \,,
\end{eqnarray}
where $\pmb{q}$ represent the momenta of incoming photon, and $\pmb{p}$ and $\pmb{p}'$ 
denote the momenta of the incoming and outgoing leptons, respectively. 
Since there are $3!$ ways of arranging the 
interactions in $\mathcal{H}$ to 
generate the same matrix element we have   
\begin{eqnarray}
\langle \pmb{p}'|T\Big((-i)^3 \int d^4xg_iX_i\overbar{\psi^c}P_{\xi}\psi\int d^4y g_i^* X_i^* \overbar{\psi}P_{\xi'}\psi^c
\int d^4z (-eQ)\overbar{\psi}\gamma^{\mu}\psi A_{\mu}\Big)|\pmb{p}\ \pmb{q}\rangle\,. \label{smatrix}
\end{eqnarray}
There are four different ways of contracting the fields in Eq.~(\ref{smatrix}): 
\begin{eqnarray}
\langle\contraction{}{\pmb{p}'}{|\int d^4xX_i\overbar{\psi_a^c}(P_{\xi})_{aa'}\psi_{a'}\int d^4yX_i^*}{\overbar{\psi_b}}
\bcontraction[2ex]{\pmb{p}'|\int d^4x}{X_i}{\overbar{\psi_a^c}(P_{\xi})_{aa'}\psi_{a'}\int d^4y}{X_i^*}
\contraction[2ex]{\pmb{p}'|\int d^4xX_i}{\overbar{\psi_a^c}}{(P_{\xi})_{aa'}\psi_{a'}\int d^4y X_i^* \overbar{\psi_b}
(P_{\xi'})_{bb'}\psi^c_{b'}\int d^4z}{\overbar{\psi_c}}
\contraction[2ex]{\pmb{p}'|\int d^4xX_i\overbar{\psi_a^c}(P_{\xi})_{aa'}}{\psi_{a'}}{\int d^4y X_i^* \overbar{\psi_b}
(P_{\xi'})_{bb'}\psi^c_{b'}\int d^4z \overbar{\psi_c}\gamma^{\mu}_{cd}\psi_d A_{\mu}|}{\pmb{p}}
\bcontraction[2ex]{\pmb{p}'|\int d^4xX_i\overbar{\psi_a^c}(P_{\xi})_{aa'}\psi_{a'}\int d^4y X_i^* \overbar{\psi_b}
(P_{\xi'})_{bb'}}{\psi^c_{b'}}{\int d^4z \overbar{\psi_c}\gamma^{\mu}_{cd}}{\psi_d}
\bcontraction[2ex]{\pmb{p}'|\int d^4xX_i\overbar{\psi_a^c}(P_{\xi})_{aa'}\psi_{a'}\int d^4y X_i^* \overbar{\psi_b}
(P_{\xi'})_{bb'}\psi^c_{b'}\int d^4z \overbar{\psi_c}\gamma^{\mu}_{cd}\psi_d }{A_{\mu}}{|\pmb{p}\ }{\pmb{q}}
\pmb{p}'|\int d^4xX_i\overbar{\psi_a^c}(P_{\xi})_{aa'}\psi_{a'}\int d^4y X_i^* \overbar{\psi_b}(P_{\xi'})_{bb'}}{\psi^c_{b'}
\int d^4z \overbar{\psi_c}\gamma^{\mu}_{cd}\psi_d A_{\mu}|\pmb{p}\ \pmb{q}\rangle \,, \label{Contraction1}
\end{eqnarray}
\begin{eqnarray}
\langle\contraction{}{\pmb{p}'}{|\int d^4xX_i\overbar{\psi_a^c}(P_{\xi})_{aa'}\psi_{a'}\int d^4yX_i^*\overbar{\psi_b}(P_{\xi'})_{bb'}}{\psi^c_{b'}}
\bcontraction[2ex]{\pmb{p}'|\int d^4x}{X_i}{\overbar{\psi_a^c}(P_{\xi})_{aa'}\psi_{a'}\int d^4y}{X_i^*}
\contraction[2ex]{\pmb{p}'|\int d^4xX_i}{\overbar{\psi_a^c}}{(P_{\xi})_{aa'}\psi_{a'}\int d^4y X_i^* \overbar{\psi_b}
(P_{\xi'})_{bb'}\psi^c_{b'}\int d^4z\overbar{\psi_c}\gamma^{\mu}_{cd}\psi_d A_{\mu}|}{\pmb{p}}
\contraction[2ex]{\pmb{p}'|\int d^4xX_i\overbar{\psi_a^c}(P_{\xi})_{aa'}}{\psi_{a'}}{\int d^4y X_i^* \overbar{\psi_b}
(P_{\xi'})_{bb'}\psi^c_{b'}\int d^4z}{\overbar{\psi_c}}
\bcontraction[2ex]{\pmb{p}'|\int d^4xX_i\overbar{\psi_a^c}(P_{\xi})_{aa'}\psi_{a'}\int d^4y X_i^*}{\overbar{\psi_b}}
{(P_{\xi'})_{bb'}\psi^c_{b'}\int d^4z \overbar{\psi_c}\gamma^{\mu}_{cd}}{\psi_d}
\bcontraction[2ex]{\pmb{p}'|\int d^4xX_i\overbar{\psi_a^c}(P_{\xi})_{aa'}\psi_{a'}\int d^4y X_i^* \overbar{\psi_b}
(P_{\xi'})_{bb'}\psi^c_{b'}\int d^4z \overbar{\psi_c}\gamma^{\mu}_{cd}\psi_d }{A_{\mu}}{|\pmb{p}\ }{\pmb{q}}
\pmb{p}'|\int d^4xX_i\overbar{\psi_a^c}(P_{\xi})_{aa'}\psi_{a'}\int d^4y X_i^* \overbar{\psi_b}(P_{\xi'})_{bb'}}{\psi^c_{b'}
\int d^4z \overbar{\psi_c}\gamma^{\mu}_{cd}\psi_d A_{\mu}|\pmb{p}\ \pmb{q}\rangle \,, \label{Contraction2}
\end{eqnarray}
\begin{eqnarray}
\langle\contraction{}{\pmb{p}'}{|\int d^4xX_i\overbar{\psi_a^c}(P_{\xi})_{aa'}\psi_{a'}\int d^4yX_i^*}{\overbar{\psi_b}}
\bcontraction[2ex]{\pmb{p}'|\int d^4x}{X_i}{\overbar{\psi_a^c}(P_{\xi})_{aa'}\psi_{a'}\int d^4y}{X_i^*}
\contraction[2ex]{\pmb{p}'|\int d^4xX_i}{\overbar{\psi_a^c}}{(P_{\xi})_{aa'}\psi_{a'}\int d^4y X_i^* \overbar{\psi_b}
(P_{\xi'})_{bb'}\psi^c_{b'}\int d^4z\overbar{\psi_c}\gamma^{\mu}_{cd}\psi_d A_{\mu}}{\pmb{p}}
\contraction[2ex]{\pmb{p}'|\int d^4xX_i\overbar{\psi_a^c}(P_{\xi})_{aa'}}{\psi_{a'}}{\int d^4y X_i^* \overbar{\psi_b}
(P_{\xi'})_{bb'}\psi^c_{b'}\int d^4z}{\overbar{\psi_c}}
\bcontraction[2ex]{\pmb{p}'|\int d^4xX_i\overbar{\psi_a^c}(P_{\xi})_{aa'}\psi_{a'}\int d^4y X_i^* \overbar{\psi_b}
(P_{\xi'})_{bb'}}{\psi^c_{b'}}{\int d^4z \overbar{\psi_c}\gamma^{\mu}_{cd}}{\psi_d}
\bcontraction[2ex]{\pmb{p}'|\int d^4xX_i\overbar{\psi_a^c}(P_{\xi})_{aa'}\psi_{a'}\int d^4y X_i^* \overbar{\psi_b}
(P_{\xi'})_{bb'}\psi^c_{b'}\int d^4z \overbar{\psi_c}\gamma^{\mu}_{cd}\psi_d }{A_{\mu}}{|\pmb{p}\ }{\pmb{q}}
\pmb{p}'|\int d^4xX_i\overbar{\psi_a^c}(P_{\xi})_{aa'}\psi_{a'}\int d^4y X_i^* \overbar{\psi_b}(P_{\xi'})_{bb'}}{\psi^c_{b'}
\int d^4z \overbar{\psi_c}\gamma^{\mu}_{cd}\psi_d A_{\mu}|\pmb{p}\ \pmb{q}\rangle \,, \label{Contraction3}
\end{eqnarray}
\begin{eqnarray}
\langle\contraction{}{\pmb{p}'}{|\int d^4xX_i\overbar{\psi_a^c}(P_{\xi})_{aa'}\psi_{a'}\int d^4yX_i^*\overbar{\psi_b}(P_{\xi'})_{bb'}}{\psi^c_{b'}}
\bcontraction[2ex]{\pmb{p}'|\int d^4x}{X_i}{\overbar{\psi_a^c}(P_{\xi})_{aa'}\psi_{a'}\int d^4y}{X_i^*}
\contraction[2ex]{\pmb{p}'|\int d^4xX_i}{\overbar{\psi_a^c}}{(P_{\xi})_{aa'}\psi_{a'}\int d^4y X_i^* \overbar{\psi_b}
(P_{\xi'})_{bb'}\psi^c_{b'}\int d^4z}{\overbar{\psi_c}}
\contraction[2ex]{\pmb{p}'|\int d^4xX_i\overbar{\psi_a^c}(P_{\xi})_{aa'}}{\psi_{a'}}{\int d^4y X_i^* \overbar{\psi_b}
(P_{\xi'})_{bb'}\psi^c_{b'}\int d^4z \overbar{\psi_c}\gamma^{\mu}_{cd}\psi_d A_{\mu}|}{\pmb{p}}
\bcontraction[2ex]{\pmb{p}'|\int d^4xX_i\overbar{\psi_a^c}(P_{\xi})_{aa'}\psi_{a'}\int d^4y X_i^*}{\overbar{\psi_b}}
{(P_{\xi'})_{bb'}\psi^c_{b'}\int d^4z \overbar{\psi_c}\gamma^{\mu}_{cd}}{\psi_d}
\bcontraction[2ex]{\pmb{p}'|\int d^4xX_i\overbar{\psi_a^c}(P_{\xi})_{aa'}\psi_{a'}\int d^4y X_i^* \overbar{\psi_b}
(P_{\xi'})_{bb'}\psi^c_{b'}\int d^4z \overbar{\psi_c}\gamma^{\mu}_{cd}\psi_d }{A_{\mu}}{|\pmb{p}\ }{\pmb{q}}
\pmb{p}'|\int d^4xX_i\overbar{\psi_a^c}(P_{\xi})_{aa'}\psi_{a'}\int d^4y X_i^* \overbar{\psi_b}(P_{\xi'})_{bb'}}{\psi^c_{b'}
\int d^4z \overbar{\psi_c}\gamma^{\mu}_{cd}\psi_d A_{\mu}|\pmb{p}\ \pmb{q}\rangle \,, \label{Contraction4}
\end{eqnarray}
where we have factored out $-(-i)^3g_ig^*_iQe$ and have left the 
spinor indices explicit. After some manipulation
we find each contribution is identical, so that 
after pulling out the factor $(2\pi)^4\delta^4(p+q-p')$, 
the total matrix element is 
\begin{eqnarray}
i \mathcal{M}^{\mu}= 4Qe|g_i|^2\int\frac{d^4k}{(2\pi)^4}\frac{\bar{u}(p')P_{\xi'}(\slashed{k}-m_b)\gamma^{\mu}(\slashed{k}'-m_b)P_{\xi}u(p)}{(k^2-m^2_b + i\epsilon)(k'^2-m^2_b +i\epsilon)((k+p')^2-M^2_{X_i} +i\epsilon )}\,, \label{amplitude01}
\end{eqnarray}
where  $k'=k+q$ and $m_b$ and $M_{X_i}$ are the masses of the charged 
lepton and scalar in the loop, respectively --- 
 the overall 4 comes from the different contractions we have noted. 
We find that Eq.~(\ref{amplitude01}) contributes to $a_{\ell_a}$ as 
\begin{eqnarray}
\delta a_{\ell_a}
=\frac{Qg_ig^*_i}{4\pi^2}\int^1_0 dz\frac{m^2_az(1-z)^2}{(z^2-z)m^2_a+zM^2_{X_i}+(1-z)m^2_b}\,,\label{g2a}
\end{eqnarray} 
where $m_a$ is the mass of external lepton a. 
Note that the final result is independent of $\xi$. 

We now move to the second case. The interaction is 
\begin{eqnarray}
\mathcal{H} \supset  -iQ'e[(\partial^{\mu}X_i)X_i^{*}-X_i(\partial^{\mu}X_i^*)]A_{\mu} +
g_iX_i\overbar{\psi^c}P_{\xi}\psi + g_i^* X_i^* \overbar{\psi}P_{\xi'}\psi^c \,,
\end{eqnarray}
where the charged scalar has $Q'=2$, if it couples to two electrons. 
Here, too, 
there are four different contractions, and they 
contribute identically to the one-loop amplitude.  Since 
there is only one way to contract all the scalars, we 
show it separately from the four different fermion
contractions: 
\begin{eqnarray}
&&\langle\contraction{}{\pmb{p}'}{|\int d^4x\overbar{\psi_a^c}(P_{\xi})_{aa'}\psi_{a'}\int d^4y}{\overbar{\psi_b}}
\bcontraction[2ex]{\pmb{p}'|\int d^4x}{\overbar{\psi_a^c}}{(P_{\xi})_{aa'}\psi_{a'}\int d^4y  \overbar{\psi_b}
(P_{\xi'})_{bb'}}{\psi^c_{b'}}
\contraction[2ex]{\pmb{p}'|\int d^4x\overbar{\psi_a^c}(P_{\xi})_{aa'}}{\psi_{a'}}{\int d^4y \overbar{\psi_b}
(P_{\xi'})_{bb'}\psi^c_{b'}\int d^4zA_{\mu}}{|\pmb{p}}
\bcontraction{\pmb{p}'|\int d^4x\overbar{\psi_a^c}(P_{\xi})_{aa'}\psi_{a'}\int d^4y \overbar{\psi_b}
(P_{\xi'})_{bb'}\psi^c_{b'}\int d^4z}{A_{\mu}}{|\pmb{p}\ }{\pmb{q}}
\pmb{p}'|\int d^4x\overbar{\psi_a^c}(P_{\xi})_{aa'}\psi_{a'}\int d^4y \overbar{\psi_b}(P_{\xi'})_{bb'}}{\psi^c_{b'}
\int d^4z A_{\mu}|\pmb{p}\ \pmb{q}\rangle \,, \\   \label{Contraction1b} 
&&\langle\contraction{}{\pmb{p}'}{|\int d^4x\overbar{\psi_a^c}(P_{\xi})_{aa'}\psi_{a'}\int d^4y \overbar{\psi_b}(P_{\xi'})_{bb'}}{\psi^c_{b'}}
\contraction[2ex]{\pmb{p}'|\int d^4x}{\overbar{\psi_a^c}}{(P_{\xi})_{aa'}\psi_{a'}\int d^4y\overbar{\psi_b}(P_{\xi'})_{bb'}\psi^c_{b'}\int d^4z A_{\mu}|}{\pmb{p}}
\bcontraction[2ex]{\pmb{p}'|\int d^4x\overbar{\psi_a^c}(P_{\xi})_{aa'}}{\psi_{a'}}{\int d^4y}{\overbar{\psi_b}}
\bcontraction{\pmb{p}'|\int d^4x\overbar{\psi_a^c}(P_{\xi})_{aa'}\psi_{a'}\int d^4y \overbar{\psi_b}
(P_{\xi'})_{bb'}\psi^c_{b'}\int d^4z}{A_{\mu}}{|\pmb{p}\ }{\pmb{q}}
\pmb{p}'|\int d^4x\overbar{\psi_a^c}(P_{\xi})_{aa'}\psi_{a'}\int d^4y \overbar{\psi_b}(P_{\xi'})_{bb'}}{\psi^c_{b'}
\int d^4z A_{\mu}|\pmb{p}\ \pmb{q}\rangle \,, \\  \label{Contraction2b}
&&\langle\contraction{}{\pmb{p}'}{|\int d^4x\overbar{\psi_a^c}(P_{\xi})_{aa'}\psi_{a'}\int d^4y}{\overbar{\psi_b}}
\contraction[2ex]{\pmb{p}'|\int d^4x}{\overbar{\psi_a^c}}{(P_{\xi})_{aa'}\psi_{a'}\int d^4y  \overbar{\psi_b}
(P_{\xi'})_{bb'}\psi^c_{b'}\int d^4z A_{\mu}|}{\pmb{p}}
\bcontraction[2ex]{\pmb{p}'|\int d^4x\overbar{\psi_a^c}(P_{\xi})_{aa'}}{\psi_{a'}}{\int d^4y \overbar{\psi_b}
(P_{\xi'})_{bb'}}{\psi^c_{b'}}
\bcontraction{\pmb{p}'|\int d^4x\overbar{\psi_a^c}(P_{\xi})_{aa'}\psi_{a'}\int d^4y \overbar{\psi_b}
(P_{\xi'})_{bb'}\psi^c_{b'}\int d^4z}{A_{\mu}}{|\pmb{p}\ }{\pmb{q}}
\pmb{p}'|\int d^4x\overbar{\psi_a^c}(P_{\xi})_{aa'}\psi_{a'}\int d^4y \overbar{\psi_b}(P_{\xi'})_{bb'}}{\psi^c_{b'}
\int d^4z A_{\mu}|\pmb{p}\ \pmb{q}\rangle \,, \\  \label{Contraction3b}
&&\langle\contraction{}{\pmb{p}'}{|\int d^4x\overbar{\psi_a^c}(P_{\xi})_{aa'}\psi_{a'}\int d^4y \overbar{\psi_b}(P_{\xi'})_{bb'}}{\psi^c_{b'}}
\bcontraction[2ex]{\pmb{p}'|\int d^4x}{\overbar{\psi_a^c}}{(P_{\xi})_{aa'}\psi_{a'}\int d^4y}{\overbar{\psi_b}}
\contraction[2ex]{\pmb{p}'|\int d^4x\overbar{\psi_a^c}(P_{\xi})_{aa'}}{\psi_{a'}}{\int d^4y \overbar{\psi_b}
(P_{\xi'})_{bb'}\psi^c_{b'}\int d^4z A_{\mu}|}{\pmb{p}}
\bcontraction{\pmb{p}'|\int d^4x\overbar{\psi_a^c}(P_{\xi})_{aa'}\psi_{a'}\int d^4y \overbar{\psi_b}
(P_{\xi'})_{bb'}\psi^c_{b'}\int d^4z}{A_{\mu}}{|\pmb{p}\ }{\pmb{q}}
\pmb{p}'|\int d^4x\overbar{\psi_a^c}(P_{\xi})_{aa'}\psi_{a'}\int d^4y \overbar{\psi_b}(P_{\xi'})_{bb'}}{\psi^c_{b'}
\int d^4z A_{\mu}|\pmb{p}\ \pmb{q}\rangle \,, \label{Contraction4b} 
\end{eqnarray}
with 
\begin{eqnarray}
ig_ig_i^*Q'e\Big[\contraction{}{X}{{}_i(x)X^*_i(y)(i\partial^{\mu}X_i(z))}{X}
\contraction[2ex]{X_i(x)}{X}{{}^*_i(y)(i\partial^{\mu}}{X}
\contraction{X_i(x)X^*_i(y)(i\partial^{\mu}X_i(z))X_i^{*}(z)-}{X}{{}_i(x)X^*_i(y)X_i(z)(i\partial^{\mu}}{X}
\contraction[2ex]{X_i(x)X^*_i(y)(i\partial^{\mu}X_i(z))X_i^{*}(z)-X_i(x)}{X}{{}^*_i(y)}{X}
X_i(x)X^*_i(y)(i\partial^{\mu}X_i(z))X_i^{*}(z)-X_i(x)X^*_i(y)X_i(z)(i\partial^{\mu}X_i^*(z))\Big]\,.
\end{eqnarray}
After combining all of the contractions and dropping the 
factor $(2\pi)^4\delta^4(p+q-p')$, we find the one-loop matrix element 
in the second case is 
\begin{eqnarray}
i\mathcal{M}^{\mu}=-4g_ig^*_iQ'e\int\frac{d^4k}{(2\pi)^4}\frac{\bar{u}(p')P_{\xi'}(\slashed{k}+\slashed{p}'+m_b)u(p)(k+k')^{\mu}}{(k^2-M^2_{X_i})(k'^2-M^2_{X_i})((k+p')^2-m^2_b)}\,, \label{amplitude02}
\end{eqnarray}
with $k'=k+q$, which contributes to $a_{\ell_a}$ as 
\begin{eqnarray}
\delta a_{\ell_a}
=\frac{-Q'g_ig^*_i}{4\pi^2}\int^1_0 dz\frac{m^2_az(1-z)^2}{(z^2-z)m^2_a+zm^2_b+(1-z)M^2_{X_i}}\,,\label{g2b}
\end{eqnarray} 
noting that this result is independent of $\xi$, too.  
To compute the final contribution to $\delta a_{\ell_a}$ from either $X_1$ or 
$X_3^3$ we add those of Eqs.~(\ref{g2a},\ref{g2b}). 

The computation of $\delta a_{\ell_e}$ from $X_2$, or from $X_3^2$, 
 is more straightforward in that only a single set of fermion contractions 
exists. We find from $X_2$, where $Q'=1$ for the scalar that couples to an electron 
and a neutrino, that 
\begin{eqnarray}
\delta a_{\ell_a}
=\frac{-Q'4 |g_2^{1b}|^2}{16\pi^2}\int^1_0 dz\frac{m^2_az(1-z)^2}{(z^2-z)m^2_a+zm^2_b+(1-z)M^2_{X_i}}\,,\label{g2c}
\end{eqnarray} 
where $m_b^2 = m_{\nu_b}^2$. To find 
the contribution from $X_3^2$ we replace $2 g_2^{1b}$ with $\sqrt{2}g_3^{11}$ and
note that $m_b^2$ is just $m_{\nu_e}^2$. 

\bibliography{gm2epuzz_final}

\end{document}